\documentclass[pdflatex,sn-mathphys-num,iicol]{sn-jnl}
\usepackage[T1]{fontenc}
\usepackage{amsmath,amssymb,bm,mathrsfs}
\DeclareFontFamily{U}{rsfs}{\skewchar\font127}
\DeclareFontShape{U}{rsfs}{m}{n}{
  <-7> rsfs5
  <7-8> rsfs7
  <8-> rsfs10
}{}
\usepackage{graphicx}
\usepackage{booktabs}
\usepackage{array}
\usepackage{microtype}
\hypersetup{hidelinks}
\newcommand{\mus}{\mu_{\rm s}}
\newcommand{\om}{\widehat{\omega}}
\newcommand{\muh}{\widehat{\mu}}
\newcommand{\Gam}{\widehat{\Gamma}}
\newcommand{\RN}{Reissner--Nordstr{\"o}m}

\begin{document}
\title[Lifetime reversals of scalar clouds around nonminimal EYM black holes]
{Lifetime reversals of neutral and charged scalar clouds around
nonminimal Einstein--Yang--Mills black holes}

\author*[1]{\fnm{S. V.} \sur{Bolokhov}}
\email{bolokhov-sv@rudn.ru}

\affil*[1]{\orgname{RUDN University},
\orgaddress{\street{6 Miklukho-Maklaya St}, \city{Moscow},
\postcode{117198}, \country{Russian Federation}}}

\abstract{
{We calculate quasi-bound states of neutral gauge-singlet scalars
and charged scalar doublets and triplets around the asymptotically flat,
nonminimal Einstein--Yang--Mills black hole.
For neutral monopoles, the principal result is a reversal
in the lifetime response to the nonminimal coupling:}
weakly bound clouds decay more slowly than on the corresponding
Reissner--Nordstr{\"o}m geometry, whereas increasing the field mass changes
suppression into enhancement of absorption. The weak-binding behavior
approaches a horizon-area law, but that law alone does not determine the
finite-mass ordering. At representative couplings, dipole and quadrupole
widths instead decrease substantially while their level energies change
very little. Two-sided integration and an independent arbitrary-precision
continued fraction check both the real frequencies and the much smaller
imaginary parts. We recover the Schwarzschild and
Reissner--Nordstr{\"o}m limits, test hydrogenic binding and the low-frequency
absorption scaling, and distinguish the loss of a potential well from the
disappearance of a quasi-bound pole. {An exact horizon-domain
analysis fixes the black-hole parameter range, and a positive-potential
argument excludes exponentially growing modes of the canonical test scalars.}
{The magnetic Wu--Yang connection separates the charged
multiplets into quantized monopole sectors, without an electric frequency
shift. Their lowest channels exhibit twelve located lifetime crossings
within the sampled trapping region and noninteger weak-binding decay
powers. At fixed total angular momentum within the triplet, charged
states can instead decay hundreds of times faster than neutral states.
These comparisons show that neither gauge charge nor nonminimal
coupling universally prolongs cloud survival.}
}


\maketitle

\section{Introduction}
\label{sec:intro}

A black-hole horizon absorbs waves, but a massive field need not disappear
on the light-crossing timescale. Below the rest-mass threshold, its distant
tail can decrease exponentially. An exterior potential well then supports
localized resonances whose energies resemble atomic levels and whose finite
lifetimes measure leakage into the horizon. These quasi-bound states have
been studied since the early analyses of relativistic gravitational binding
\cite{DeruelleRuffini:1974,DamourDeruelleRuffini:1976,Detweiler:1980}.
Continued-fraction calculations and time-dependent studies established
their spectra and the persistence of massive scalar configurations around
Schwarzschild and Kerr black holes
\cite{Dolan:2007,GrainBarrau:2008,Barranco:2012}. The gravitational-atom
description emphasizes that binding and absorption sample different radial
regions \cite{Baumann:2019}.

The nonminimal Einstein--Yang--Mills (EYM) black hole provides a
particularly useful setting for separating these effects. Balakin and
Zayats constructed its asymptotically flat geometry from a curvature-coupled
Wu--Yang gauge field \cite{Balakin:2007}; Balakin, Lemos, and Zayats extended
the analysis to a cosmological constant and classified the horizons
\cite{Balakin:2016}. A magnetic charge changes the asymptotic
\(1/r^2\) term, whereas the nonminimal correction to this particular
metric starts at \(1/r^5\). The two parameters thus affect the cloud's
distant binding region and its inner absorption barrier differently.

The radiative spectrum is already known to be sensitive to this background.
L{\"u}tf{\"u}o{\u g}lu computed massless-scalar quasinormal modes (QNMs)
and grey-body factors using continued fractions and semiclassical checks,
finding much larger changes in QNM overtones than in the fundamental mode
\cite{Lutfuoglu:2025ljm}. Dubinsky subsequently studied massive-field
resonances in asymptotically flat and de Sitter members of the same family
\cite{Dubinsky:2025EYM}.

{More broadly, massive-field QNMs have been studied extensively
on static, charged, and rotating backgrounds
\cite{Ohashi:2004wr,Konoplya:2018qov,Konoplya:2017tvu,
Percival:2020skc,Kokkotas:2010zd,Konoplya:2002wt}, regular and quantum-corrected geometries
\cite{Bolokhov:2023ruj,Skvortsova:2024eqi,Skvortsova:2025cah,
Skvortsova:2026jtx,Lutfuoglu:2025hwh,Bolokhov:2024bke,
Lutfuoglu:2026gis,Lutfuoglu:2026fpx,Dubinsky:2026nww}, and
modified-gravity, environmentally deformed, and brane-world spacetimes
\cite{Lutfuoglu:2025hjy,Lutfuoglu:2025kqp,Lutfuoglu:2025qkt,
Lutfuoglu:2025bsf,Bolokhov:2026dzn,Bolokhov:2026dfg,
Skvortsova:2026unq,Dubinsky:2025EYM,Lutfuoglu:2026xlo}.
Related investigations address late-time tails
\cite{Bolokhov:2024ixe,Konoplya:2023fmh,Konoplya:2006gq,
Dubinsky:2024hmn} and grey-body factors, absorption, and thermodynamics
\cite{Lutfuoglu:2025eik,Lutfuoglu:2026pgn}.
By comparison, systematic quasi-bound-state calculations cover a more
limited range of geometries, including Schwarzschild
\cite{GrainBarrau:2008,Barranco:2012,RosaDolan:2012}, Kerr \cite{Dolan:2007},
Reissner--Nordstr{\"o}m \cite{Sampaio:2014swa}, Kerr--Newman
\cite{Furuhashi:2004jk}, and selected regular black holes such as the
Ay\'on-Beato--Garc\'ia solution \cite{Dolan:2024RegularQBS}.}

{Here we focus on the quasi-bound
branches connected to the weak-binding hydrogenic spectrum, examining
how the nonminimal geometry changes the binding and decay of exterior
scalar clouds.}

Our main question is whether increasing the nonminimal coupling necessarily
protects a cloud from absorption. The answer is mass and angular-momentum
dependent. We resolve a transition from reduced to enhanced monopole
absorption, while the higher-multipole reference states exhibit substantial
suppression. This is more informative than a comparison of a single
frequency near the edge of the black-hole domain. We therefore give spectra
at several field masses and several nonminimal couplings, distinguish
fixed-coupling charge scans from scans at fixed fraction of extremality,
and compare the results with the weak-binding horizon-area limit.

{We first establish the neutral gauge-singlet spectrum and then
calculate canonical charged SU(2) scalar doublets and triplets. Because
the source is magnetic, this extension requires a gauge-covariant angular
problem. The reducible
Wu--Yang connection separates into quantized monopole sectors
\cite{Balakin:2007,WuYang:1976}, whose charged spectra are presented in
Sec.~\ref{sec:charged}. The new comparison distinguishes nonminimal
lifetime reversals within a charged sector from enhanced absorption at
fixed total angular momentum within the triplet.}

The paper proceeds from the exact horizon domain and a
{test-scalar stability argument, first for the neutral field
and then for the canonical charged sectors,} to two numerical constructions. One integrates the
radial equation from both boundaries; the other imposes an arbitrary-
precision continued fraction. The comparison tests the imaginary parts
separately, since agreement in the real frequency alone cannot validate
a very narrow decay width. The resulting spectral data determine the quantitative conclusions.

\section{Geometry and parameter domain}
\label{sec:geometry}

\subsection{Background and conventions}

We use \(G=c=\hbar=1\) and the metric
\begin{align}
 ds^2&=-f(r)dt^2+f(r)^{-1}dr^2+r^2d\Omega^2,\\
 f(r)&=1+\frac{r^4}{r^4+2\xi Q^2}
       \left(\frac{Q^2}{r^2}-\frac{2M}{r}\right).
 \label{eq:metric}
\end{align}
Here \(M\) is the Arnowitt--Deser--Misner {(ADM)} mass, \(Q\) is the magnetic
charge parameter, and \(\xi\geq0\) has dimensions of length squared.
{The ADM mass is the total mass measured at spatial infinity;
\(r\) is the areal radius, \(d\Omega^2=d\theta^2+\sin^2\theta\,d\phi^2\),
and \(f\) is the lapse function. ``Nonminimal'' refers here to an explicit
curvature coupling of the background gauge field; the test scalar is minimally coupled.}
The underlying nonminimal EYM theory contains a susceptibility tensor
contracted with two Yang--Mills field strengths. In the conventions of
Refs.~\cite{Balakin:2007,Balakin:2016}, it is
\begin{align}
 \mathcal R^{ikmn}={}&\frac{q_1R}{2}
 (g^{im}g^{kn}-g^{in}g^{km})\nonumber\\
 &+\frac{q_2}{2}(R^{im}g^{kn}-R^{in}g^{km}
 +R^{kn}g^{im}-R^{km}g^{in})\nonumber\\
 &+q_3R^{ikmn},
 \label{eq:susceptibility}
\end{align}
with \((q_1,q_2,q_3)=(-\xi,4\xi,-6\xi)\) on the regular one-parameter
branch. This coupling combines Ricci-scalar, Ricci-tensor, and Riemann-tensor contributions.

{In this expression, \(R\), \(R^{ik}\), and \(R^{ikmn}\) are the
Ricci scalar, Ricci tensor, and Riemann tensor; \(g^{ik}\) is the inverse
metric. The susceptibility \(\mathcal R^{ikmn}\) weights the response
of the Yang--Mills field strength to curvature. Spacetime indices run over
four coordinates, and the parenthesized index in \(A_\mu^{(a)}\) labels
the three internal SU(2) gauge directions.} The Wu--Yang solution is purely
magnetic, with \(A_t^{(a)}=0\). We use its metric as a prescribed
background and do not perturb the coupled metric--Yang--Mills system.

At \(\xi=0\), Eq.~\eqref{eq:metric} is the  \RN\ {(RN)} metric;
for a neutral scalar its radial equation is identical to that on an
electrically charged \RN\ geometry with the same \(Q^2\).
At \(Q=0\) and finite \(\xi\), the geometry is exactly Schwarzschild.
For \(\xi Q^2>0\), the central expansion is
\begin{equation}
 f(r)=1+\frac{r^2}{2\xi}-\frac{Mr^3}{\xi Q^2}+O(r^6).
\end{equation}
The spacetime curvature is finite at the center, although regularity of
the metric does not imply regularity of the gauge-field invariant
\cite{Balakin:2007}. No boundary condition at that center enters the
exterior eigenvalue problem.

{The scalar rest mass is \(\mus>0\), and \(\omega\) is its
possibly complex mode frequency measured with respect to the time \(t\).}
For numerical work we introduce
\begin{equation}
 x=\frac rM,\quad e=\frac{|Q|}{M},\quad b=\frac{\xi}{M^2},
 \quad\muh=M\mus,\quad\om=M\omega.
 \label{eq:units}
\end{equation}
The dimensionless lapse is
\begin{align}
 f(x)&=\frac{N(x)}{D(x)},\qquad D=x^4+2be^2,\nonumber\\
 N(x)&=x^4-2x^3+e^2x^2+2be^2.
 \label{eq:polynomials}
\end{align}
Its large-radius expansion follows directly:
\begin{equation}
 f(x)=1-\frac2x+\frac{e^2}{x^2}
       +\frac{4be^2}{x^5}-\frac{2be^4}{x^6}+O(x^{-9}).
 \label{eq:farMetric}
\end{equation}
Thus the nonminimal correction falls off faster than the charge term.

\subsection{Horizons and fixed quantities}

For \(0<e<1\), the double-horizon curve is obtained from \(N=N'=0\):
\begin{equation}
 x_{\rm ext}=\frac{3+\sqrt{9-8e^2}}4,\qquad
 b_{\rm ext}(e)=\frac{x_{\rm ext}^3(x_{\rm ext}-1)}{2e^2}.
 \label{eq:extremal}
\end{equation}
There are two simple positive horizons for \(0<b<b_{\rm ext}\),
which coalesce at equality. For larger \(b\), no black-hole horizon
exists. One way to obtain the full domain without relying on a quartic
root classifier is to observe that, for \(b>0\), a positive root requires
\(x^2-2x+e^2<0\), so \(e<1\), and that increasing the constant
term \(2be^2\) lifts the unique relevant minimum of \(N\).
The \(b=0\) endpoints recover
\(x_\pm=1\pm\sqrt{1-e^2}\).

{For a two-horizon geometry, we denote the outer event horizon by \(x_+=r_+/M\) and the inner
Cauchy horizon by \(x_-=r_-/M\). The latter is an inner boundary of
predictability; the exterior wave calculation begins at the event horizon.
Extremality means that the two horizons coalesce into a double root;
a nonextremal horizon is a simple root. Subscripts \(x\) denote
\(d/dx\), and \(N'\) also denotes an \(x\)-derivative here.}

We parameterize the nonextremal survey by
\begin{equation}
 \lambda=\frac{b}{b_{\rm ext}(e)},\qquad 0\leq\lambda\leq0.9.
 \label{eq:lambda}
\end{equation}
The exactly extremal problem is excluded. Holding \(e\) fixed, both
\(\lambda\) and \(b\) increase together and the outer horizon decreases:
\begin{equation}
 \left.\frac{dx_+}{db}\right|_e=-\frac{2e^2}{N'(x_+)}<0.
 \label{eq:horizonDerivative}
\end{equation}
{Each ratio between geometries is formed at fixed \(M\)
and \(\muh\); mass-response curves then vary \(\muh\) from one such
comparison to the next. The additional gauge-normalization restrictions
on cross-charge comparisons are stated in Sec.~\ref{sec:charged}.}
A scan at fixed \(\lambda\) varies \(b\) through the charge dependence
of \(b_{\rm ext}\). In particular,
\(b_{\rm ext}\sim27/(32e^2)\) as \(e\to0\): keeping
\(\lambda>0\) fixed in this limit sends \(\xi/M^2\) to infinity
and leaves \(2be^2\to27\lambda/16\). It does not recover the
finite-coupling Schwarzschild limit. We therefore include Schwarzschild
as a separate reference and also perform a charge scan at fixed \(b\).

The earlier QNM calculation \cite{Lutfuoglu:2025ljm} uses horizon units
in which \(r_h=1\) and \(M=(1+Q^2+2\xi Q^2)/2\).
For parameters \(Q_h=Q/r_h\), \(\xi_h=\xi/r_h^2\), the conversion is
\begin{equation}
 \eta=\frac{M}{r_h}=\frac{1+Q_h^2+2\xi_hQ_h^2}{2},\quad
 e=\frac{Q_h}{\eta},\quad b=\frac{\xi_h}{\eta^2}.
\end{equation}
Likewise \(M\omega=\eta(r_h\omega)\) and
\(M\mus=\eta(r_h\mus)\). Apparent trends obtained while changing
\(\eta\) cannot be compared directly with fixed-ADM-mass trends.

\section{Neutral scalar spectrum}
\label{sec:field}

The scalar is a gauge singlet with positive mass squared and satisfies
\begin{equation}
 (\Box-\mus^2)\Phi=0,\qquad
 \Phi=e^{-i\omega t}Y_{\ell m}(\theta,\phi)\frac{u(r)}r.
\end{equation}
{For separability of the Klein--Gordon and Hamilton--Jacobi
equations in more general black-hole spacetimes, see
Refs.~\cite{Carter:1968ks,Konoplya:2018arm}.}
{A gauge singlet is invariant under the internal gauge
group and therefore has no Yang--Mills charge. Here \(\Box=\nabla_\mu
\nabla^\mu\) is the covariant wave operator, \(Y_{\ell m}\) are spherical
harmonics, and \(u\) is the radial amplitude after the geometric \(1/r\)
factor is removed. The tortoise coordinate \(r_*\), defined by
\(dr_*/dr=1/f\), sends the event horizon to \(r_*\to-\infty\).
Primes on \(f(r)\) denote \(d/dr\).}
There is no scalar curvature coupling. The radial equation is
\begin{align}
 \frac{d^2u}{dr_*^2}+[\omega^2-V_\ell(r)]u&=0,\\
 V_\ell(r)&=f(r)\left[\mus^2+\frac{\ell(\ell+1)}{r^2}
                         +\frac{f'(r)}r\right].
 \label{eq:potential}
\end{align}
The angular index \(\ell=0,1,2\) denotes the monopole, dipole, and
quadrupole, respectively; spherical symmetry removes the dependence on
\(m\). {For the neutral harmonics,
\(\ell=0,1,\ldots\) and \(m=-\ell,\ldots,\ell\); \(m\) labels the
azimuthal dependence.} The radial number \(n_r=0,1,2,\ldots\) labels continuation from
the hydrogenic ground state and its radial excitations at fixed \(\ell\).
At weak binding, the principal number
\(\mathcal N=\ell+n_r+1\) gives
\begin{equation}
 \frac{\om_R}{\muh}=1-\frac{\muh^2}{2\mathcal N^2}
                  +O(\muh^4),\qquad \muh\ll1.
 \label{eq:hydrogenic}
\end{equation}
The binding fraction therefore approaches
\(B=\muh^2/(2\mathcal N^2)\). This relation identifies the
weak-binding branch; relativistic corrections are retained by both
numerical methods.

\subsection{Boundary conditions and widths}

In dimensionless variables define
\begin{align}
 k&=\sqrt{\muh^2-\om^2},\qquad \Re k>0,\nonumber\\
 \nu&=\frac{2\om^2-\muh^2}{k},\qquad
 \rho=-\frac{i\om}{f_x(x_+)}.
 \label{eq:exponents}
\end{align}
{Here \(k\) is the complex inverse decay length in units
of \(x\), \(\rho\) is the horizon power, and \(\nu\) is the far-field
Coulomb power. A square-root sheet specifies the continuous choice of
sign of the complex square root; the condition \(\Re k>0\) selects
the exponentially decreasing spatial solution.}
The desired solutions satisfy
\begin{equation}
 u\sim(x-x_+)^\rho\quad(x\to x_+),\qquad
 u\sim e^{-kx}x^\nu\quad(x\to\infty).
 \label{eq:boundaries}
\end{equation}
The factor \(x^\nu\) is the gravitational Coulomb correction; retaining
only an exponential would give unnecessarily inaccurate finite-radius
boundary data. The positive-real-part branch of \(k\) is followed
continuously during root finding. An outgoing QNM condition instead
continues the radiative solution at infinity. A small imaginary part does
not make these two boundary conditions equivalent
\cite{Dolan:2007,KonoplyaZhidenko:2011}.

We write \(\om=\om_R+i\om_I\), {where the subscripts
\(R,I\) denote real and imaginary parts,} and define the amplitude decay width,
amplitude lifetime, and fractional binding by
\begin{equation}
 \Gam=-\om_I,\qquad \frac{\tau_{\rm amp}}M=\frac1\Gam,
 \qquad B=1-\frac{\om_R}{\muh}.
 \label{eq:observables}
\end{equation}
{When mode indices are displayed, \(\Gam_{\ell n_r}\)
labels the neutral angular and radial quantum numbers; in particular,
\(\Gam_{00}\) is the fundamental monopole width.}
The energy decreases twice as fast as the amplitude:
\(\tau_E=\tau_{\rm amp}/2\). All widths quoted below use the amplitude convention. A smaller
\(\Gam\) means a longer-lived cloud. Width ratios below are always
formed at the same \((\muh,\ell,n_r)\).

{For the time convention \(e^{-i\omega t}\), the mode amplitude is
proportional to \(e^{\omega_I t}\), and a quadratic cloud-energy measure
to \(e^{2\omega_I t}\). Thus the lifetimes are e-folding times. Positive
\(B\) measures the fractional deficit of the real level energy below the
rest-mass threshold. ``Suppression'' and ``enhancement'' refer to width
ratios below and above unity relative to the stated reference geometry;
a lifetime reversal or crossover marks a change in the relative decay rates. For charged comparisons, the fixed angular
labels are \((p,j)\), as specified in Sec.~\ref{sec:charged}.}

\subsection{Trapping and neutral-field stability}

Let \(\mathcal V=M^2V\). A local maximum followed by
a local minimum provides the geometrical setting for trapping. For a cloud
interpreted through three real turning points, the eigenvalue must also satisfy
\begin{equation}
 \mathcal V_{\min}<\om_R^2<\min(\mathcal V_{\max},\muh^2).
 \label{eq:trapping}
\end{equation}
This real-frequency criterion selects the resonances used in the
tunneling-cloud interpretation. Quasi-bound poles can continue beyond
this trapping region under the same ingoing/decaying boundary conditions.
The merger of potential extrema occurs at
\(\mathcal V_x=\mathcal V_{xx}=0\). Since
\(\mathcal V=f\muh^2+\mathcal V_0\), it can be calculated from
\begin{equation}
 \muh_{\rm well}^2=\max_{x>x_+}
             \left[-\frac{\mathcal V_{0,x}}{f_x}\right],
 \qquad \mathcal V_0=f\left[\frac{\ell(\ell+1)}{x^2}
                                      +\frac{f_x}{x}\right].
 \label{eq:wellLimit}
\end{equation}
This threshold marks the disappearance of the potential well;
quasi-bound poles may persist at larger field masses.

{The turning points are the real roots \(x_1<x_2<x_3\) of
\(\mathcal V=\om_R^2\): \((x_1,x_2)\) is the inner forbidden barrier
and \((x_2,x_3)\) the classically allowed outer well. The criterion is
only geometric; a reliable weak-leakage interpretation also requires a
small width. Since \(\mathcal V_x=f_x[\muh^2-h(x)]\), with
\(h=-\mathcal V_{0,x}/f_x\) and \(f_x>0\) as shown below, extrema
merge at a maximum of \(h\). Numerically we bracket zeros of
\(\mathcal V_x\) and \(\mathcal V-\om_R^2\) on logarithmic radial
grids and refine them with a bracketed root solver. The maximum of
\(h\) is similarly bracketed and refined by one-dimensional bounded
optimization. No complex resonance is assigned literal real classical
turning points without this stated approximation.}

The potential is positive throughout the exterior on the branch considered.
For completeness, write
\begin{equation}
 f_x=\frac{2xJ(x)}{D^2},\quad
 J=x^5-e^2x^4-6be^2x+2be^4.
\end{equation}
At a simple outer horizon, \(J(x_+)>0\), and use of \(N(x_+)=0\)
gives \(J'(x_+)=x_+^2(2x_+-e^2)(4x_+-3)>0\).
Furthermore \(J''=4x^2(5x-3e^2)>0\) for \(x\geq x_+\geq1\).
Consequently \(f_x>0\) outside the horizon and
\(\mathcal V\geq0\). The Schwarzschild endpoint has the same property.

This establishes more than the absence of growth in a finite numerical
sample. If \(\omega_I>0\), an ingoing solution decreases toward the
horizon and the bound solution decreases at infinity. Integration by
parts then gives
\begin{equation}
 \int_{-\infty}^{\infty}\left(|u_{,r_*}|^2+V|u|^2\right)dr_*
       =\omega^2\int_{-\infty}^{\infty}|u|^2dr_*.
\end{equation}
The left side is real and nonnegative. Its imaginary part requires
\(\omega_R=0\) for a growing mode; its real part then excludes the
negative value \(\omega^2=-\omega_I^2\). Thus the neutral test-scalar
problem has no exponentially growing quasi-bound mode. This proof does
not establish stability of the coupled EYM system or of its Cauchy horizon.

\section{Numerical construction and validation strategy}
\label{sec:numerics}

\subsection{Two-sided integration}

Our production calculation directly integrates
\begin{align}
 u_{xx}+P u_x+\mathcal Q u&=0,\qquad P=\frac{f_x}{f},\nonumber\\
 \mathcal Q&=\frac{\om^2}{f^2}-\frac{\muh^2}{f}
       -\frac{\ell(\ell+1)}{x^2f}-\frac{f_x}{xf}.
 \label{eq:areal}
\end{align}
An ingoing series starts at \(x_++\epsilon\); a decaying inverse-radius
series starts at \(X_\infty\). Their logarithmic derivatives are matched
at an intermediate point:
\begin{equation}
 F(\om)=\left.\frac{u_x^{\rm in}}{u^{\rm in}}\right|_{x_m}
          -\left.\frac{u_x^{\rm out}}{u^{\rm out}}\right|_{x_m}=0.
 \label{eq:matching}
\end{equation}
The normalization on each side is arbitrary. We evolve the linear pair
\((u,u_x)\), which remains regular across radial nodes. The matching point is chosen away from a node.

{The baseline neutral calculation uses ten correction
terms in each boundary series; the charged high-mass continuation uses
the modified settings given below.} Their algebraic
construction and the first corrections are given in Appendix~\ref{app:series}.
We use an eighth-order Dormand--Prince integrator with relative tolerance
\(5\times10^{-13}\), absolute tolerance \(5\times10^{-15}\),
\(\epsilon=10^{-4}\), and
\begin{equation}
 X_\infty=\max\left(100,\frac{35}{\Re k}\right),\qquad
 x_m=\max\left(10,\frac{(\ell+1)^2}{2\muh^2}\right).
 \label{eq:settings}
\end{equation}
Placing the match in the cloud region is important for weak binding:
matching too deep inside a high barrier makes the residual poorly
conditioned even when the ODE itself is accurately integrated.
Complex Newton iteration is seeded by a finite-band continued fraction
and checked against the weak-binding principal number. When a candidate
jumps to a neighboring radial level, continuation from Schwarzschild
through the minimally coupled and then the nonminimal geometry recovers
the intended branch. Small residuals alone are not accepted as evidence
of the correct radial label.

{A radial node is a
zero of the radial amplitude; linear integration crosses it without the
pole that it would introduce in the Riccati equation for \(u_x/u\).
At each boundary we may set \(u=1\) and initialize \(u_x\) with the
series logarithmic derivative. We solve \(F=0\) by complex Newton
steps \(\delta\om=F/F'\), estimating \(F'\) by a centered difference
of step \(2\times10^{-6}\muh\). Corrections larger than \(0.02\muh\)
are rejected, and a final residual above \(2\times10^{-8}\) is not
accepted; the independent width and boundary checks remain decisive.
For weakly bound levels, the diagnostic
\(\mathcal N_{\rm est}=\muh/\sqrt{2B}\) must remain within \(1/2\)
of the intended principal number. This approximate diagnostic safeguards the branch label at finite mass.}

\subsection{Arbitrary-precision continued fraction}

{The Frobenius--continued-fraction approach has a long
history in black-hole QNM calculations. Leaver developed it for Kerr
and Reissner--Nordstr{\"o}m black holes
\cite{Leaver:1985,Leaver:1990}, and Onozawa used it to study the Kerr
spectrum in detail \cite{Onozawa:1996ux}. Subsequent applications include
high overtones of Schwarzschild--de Sitter black holes
\cite{Konoplya:2004uk}, brane-localized fields on rotating backgrounds
\cite{Kanti:2006ua}, charged fields on Kerr--Newman--de Sitter
\cite{Konoplya:2007zx}, and tensor perturbations of higher-dimensional
rotating black holes \cite{Kodama:2009bf}.}

{More recent applications address higher-derivative and
quantum-corrected geometries
\cite{Konoplya:2022iyn,Bolokhov:2023dxq,Bolokhov:2023bwm,
Konoplya:2024lch,Zinhailo:2024kbq}, as well as massive-field QNMs in
brane-world models \cite{Zinhailo:2024jzt}. Benda and Matyjasek
systematically compared recurrence-based algorithms, including
continued-fraction and Hill-determinant approaches, with particular
attention to efficiency and high precision \cite{Benda:2025tni}.
For parametrized geometries, the accuracy of the metric representation
is a separate issue from convergence of the spectral calculation
\cite{Konoplya:2020hyk,Konoplya:2022iyn}.}

{Complementary developments in accurate QNM calculations
include the Laplace-transform formulation of Nollert and Schmidt
\cite{Nollert:1992ifk} and the pseudospectral studies of coupled
Kerr--Newman perturbations by Dias and collaborators
\cite{Dias:2021yju,Dias:2022oqm}; these are distinct from the
continued-fraction construction used here. Its adaptation to massive
quasi-bound states retains the horizon series and recurrence machinery
but replaces the outgoing asymptotic factor by the exponentially
decreasing solution on the bound-state sheet, as in the Kerr calculation
of Dolan \cite{Dolan:2007}. We use this established strategy to construct
an independent check of the two-sided integration.}

{The Frobenius construction expands the regular part of
the solution in a power series about the horizon, a regular singular
point of the radial equation.} For an independent construction we use
\begin{equation}
 z=\frac{x-x_+}{x-x_-},\qquad
 u=e^{-kx}(x-x_-)^\nu z^\rho\sum_{n=0}^{\infty}a_nz^n,
 \label{eq:frobenius}
\end{equation}
with \(x_-=0\) for Schwarzschild. This implements the same physical
endpoints without a finite outer boundary. After rational denominators
are cleared and the known endpoint factors are cancelled, the series
satisfies an equation with polynomial coefficients of degrees
\((13,12,11)\). It generates a generic 13-term recurrence. Gaussian
elimination reduces it to three terms and the minimal solution is selected
by a Leaver continued fraction \cite{Leaver:1985,Leaver:1990,Dolan:2007}.
Appendix~\ref{app:cf} gives the construction without listing lengthy
frequency-dependent coefficients individually.

All other finite singularities are mapped and checked before using a direct
series. A pole inside \(|z|=1\) would invalidate this implementation;
analytic continuation would then be necessary. The chosen nonextremal
validation points pass this check. The check includes the complex zeros
of the rational metric denominator as well as the real horizons.

{CF denotes continued fraction, and its depth \(N_{\rm CF}\) is the
last retained recurrence index, distinct from the principal number
\(\mathcal N\). A minimal solution is the recurrence solution whose
coefficients become negligible relative to the independent dominant
solution at large index. The fraction imposes this condition. The
singularity check maps the remaining zeros of \(N\) and \(D\), after
common-root cancellations, as well as \(x=0\); horizon and infinity
endpoint factors are treated analytically.}

Floating-point Gaussian elimination can lose digits in the extremely small
imaginary parts. We therefore use 60-decimal-digit arithmetic for the
independent validation, increasing the continued-fraction depth rather
than interpreting the root-finder residual as a truncation error.
Our reported accuracy measures are
\begin{equation}
 \Delta_\omega=|\om_{\rm CF}-\om_{\rm ODE}|,\qquad
 \delta_\Gamma=\left|\frac{\Im\om_{\rm CF}}{\Im\om_{\rm ODE}}-1\right|.
 \label{eq:errors}
\end{equation}
The second measure is indispensable: an absolute error of \(10^{-11}\)
would not be adequate for a width of \(10^{-10}\).
For the ODE check we also change the horizon offset to \(2\times10^{-4}\),
the outer multiplier from 35 to 45, the series order from 10 to 12,
the matching radius to \(\max[12,0.65(\ell+1)^2/\muh^2]\), and the
relative tolerance to \(3\times10^{-14}\).

{The absolute tolerance is always one hundredth of the relative
tolerance. We define \(\delta_\Gamma^{\rm boundary}\) by replacing
CF with the altered-boundary ODE result in Eq.~\eqref{eq:errors},
while retaining the primary ODE in the denominator. High printed
precision and a small residual do not by themselves establish a
converged imaginary part.}

\section{Spectra and numerical checks}
\label{sec:results}

\subsection{Coverage, precision, and limiting cases}

The {neutral} survey contains 331 candidate poles, including Schwarzschild,
three charges \(e=0.3,0.6,0.9\), and four nonminimal fractions
\(\lambda=0,0.3,0.6,0.9\). It combines several field masses in each
angular sector with a finer mass scan at \(e=0.6\), weak-binding checks,
and the first two radial excitations. All 331 candidates are recomputed
with the altered ODE settings in Sec.~\ref{sec:numerics}.
Six very weakly bound dipole stress tests, at \(\muh=0.025,0.035\),
are excluded from the precision-qualified spectrum: although their real
parts are stable, their extremely small widths change by as much as
\(1.1\%\) under this test. We retain these raw calculations with
explicit flags for the width uncertainty.

The remaining 325 poles have relative width changes below
\(1.4\times10^{-7}\) in the boundary sweep. Of these, 312 also
satisfy the trapping condition \eqref{eq:trapping}; the other 13 are
saved but not used as three-turning-point cloud examples. Every frequency
in the printed spectral tables passes both requirements. The precision
qualification is based on empirical numerical convergence.

\begin{table*}[tb]
\centering
\caption{Independent validation at $(\ell,\muh)=(0,0.14),(1,0.25),(2,0.35)$. $N_{\rm CF}$ is the final fraction depth at 60 decimal digits; $\Delta_\omega$ and $\delta_\Gamma$ are defined in Eq.~\eqref{eq:errors}. The last column changes all ODE boundary settings as described in the text. The reported precision is based on these comparisons.}
\label{tab:validation}
\begin{tabular}{cccrrrr}
\toprule
$e$ & $\lambda$ & $\ell$ & $N_{\rm CF}$ & $\Delta_\omega$ & $\delta_\Gamma$ & $\delta_\Gamma^{\rm boundary}$ \\
\midrule
0.0 & 0.0 & 0 & 1200 & $5.5\times10^{-17}$ & $3.0\times10^{-13}$ & $9.1\times10^{-12}$ \\
0.0 & 0.0 & 1 & 1200 & $2.8\times10^{-17}$ & $4.0\times10^{-12}$ & $1.9\times10^{-11}$ \\
0.0 & 0.0 & 2 & 1200 & $5.6\times10^{-17}$ & $2.4\times10^{-11}$ & $3.2\times10^{-11}$ \\
0.6 & 0.0 & 0 & 1200 & $5.5\times10^{-18}$ & $4.2\times10^{-14}$ & $4.4\times10^{-12}$ \\
0.6 & 0.0 & 1 & 1200 & $2.8\times10^{-17}$ & $2.7\times10^{-12}$ & $1.0\times10^{-11}$ \\
0.6 & 0.0 & 2 & 1200 & $7.0\times10^{-21}$ & $2.5\times10^{-11}$ & $1.1\times10^{-11}$ \\
0.6 & 0.6 & 0 & 1200 & $7.1\times10^{-15}$ & $4.4\times10^{-11}$ & $7.5\times10^{-13}$ \\
0.6 & 0.6 & 1 & 1200 & $2.8\times10^{-17}$ & $1.6\times10^{-11}$ & $2.9\times10^{-11}$ \\
0.6 & 0.6 & 2 & 1200 & $8.9\times10^{-21}$ & $5.9\times10^{-11}$ & $9.9\times10^{-11}$ \\
0.6 & 0.9 & 0 & 4800 & $4.0\times10^{-15}$ & $2.1\times10^{-11}$ & $2.9\times10^{-11}$ \\
0.6 & 0.9 & 1 & 4800 & $1.4\times10^{-16}$ & $2.3\times10^{-11}$ & $2.9\times10^{-11}$ \\
0.6 & 0.9 & 2 & 4800 & $5.6\times10^{-17}$ & $4.5\times10^{-10}$ & $1.3\times10^{-10}$ \\
0.9 & 0.9 & 0 & 4800 & $2.0\times10^{-15}$ & $1.8\times10^{-11}$ & $2.4\times10^{-11}$ \\
0.9 & 0.9 & 1 & 4800 & $5.6\times10^{-17}$ & $3.3\times10^{-12}$ & $9.9\times10^{-12}$ \\
0.9 & 0.9 & 2 & 4800 & $5.6\times10^{-17}$ & $6.0\times10^{-9}$ & $3.9\times10^{-9}$ \\
\bottomrule
\end{tabular}
\end{table*}

Table~\ref{tab:validation} provides a different check, using
60-digit continued fractions at 15 representative fundamental modes.
Depths 300, 600, and 1200 suffice at the less deformed points, whereas
the \(\lambda=0.9\) cases are followed through depths 2400 and 4800.
The depth dependence is important: at \(e=0.9\), \(\lambda=0.9\),
\(\ell=0\), \(\muh=0.14\), a depth-1200 result has a relative
width discrepancy of about \(1.8\times10^{-6}\), despite a root
residual below \(10^{-50}\). Increasing the depth removes this
discrepancy. The final fundamental comparisons agree within
\(10^{-12}\) in complex frequency and \(10^{-7}\) in relative
width, supporting the rounded digits reported here.
The six nonminimal radial excitations in Table~\ref{tab:excitations}
are also checked with the precision fraction. The supplied convergence
files retain the results at every depth.

There are three complementary limit checks. First, setting \(b=0\)
recovers the neutral-scalar RN radial equation exactly, while setting
\(e=0\) at finite \(b\) recovers Schwarzschild. The numerical
metric residuals in these identities are at floating-point roundoff.
Second, the Schwarzschild monopole potential-well merger is recovered
at \(x=4\), \(\muh_{\rm well}=1/4\). Third, the real spectrum
approaches the hydrogenic expression \eqref{eq:hydrogenic} as the field
mass decreases, and the monopole width approaches the independent
weak-binding scaling derived in Sec.~\ref{sec:discussion}.

\begin{table*}[tb]
\centering
\caption{Weak-binding monopole check. The central columns show $\Gam/(2x_+^2\muh^6)$, which tends to unity.
{RN uses $e=0.6$, $\lambda=0$; EYM uses $e=0.6$, $\lambda=0.9$.} The final column is the EYM/RN width ratio at $e=0.6$, $\lambda=0.9$; its limiting value is 0.716041.}
\label{tab:weak}
\begin{tabular}{crrrr}
\toprule
$\muh$ & Schwarzschild & RN & EYM & $R_{\rm RN}$ \\
\midrule
0.010 & 1.012498 & 1.012079 & 1.016766 & 0.719357 \\
0.015 & 1.026105 & 1.025282 & 1.035917 & 0.723469 \\
0.020 & 1.044010 & 1.042693 & 1.061799 & 0.729162 \\
\bottomrule
\end{tabular}
\end{table*}

{The last check tests the imaginary part against an
independently derived weak-binding limit, complementing the comparison
between the two numerical methods.} Table~\ref{tab:weak} uses additional masses as small as
\(\muh=0.01\). The quantity \(\Gam/(2x_+^2\muh^6)\) moves
toward unity in Schwarzschild, RN, and the nonminimal geometry. The
EYM/RN ratio approaches the squared horizon-radius ratio. Thus the limits are recovered in the appropriate order, without
the fixed-\(\lambda\), zero-charge ambiguity described earlier.

\subsection{Reference spectrum: energies versus absorption}

\begin{table*}[tb]
\centering
\caption{Fundamental quasi-bound frequencies in ADM-mass units. The last column is $R_{\rm S}=\Gam/\Gam_{\rm Schwarzschild}$ at the same $\ell$ and $\muh$. All listed states pass the three-turning-point test.}
\label{tab:reference}
\begin{tabular}{ccccrrr}
\toprule
$e$ & $\lambda$ & $\ell$ & $\muh$ & $\om_R$ & $\Gam=-\om_I$ & $R_{\rm S}$ \\
\midrule
0.0 & 0.0 & 0 & 0.14 & $0.13837505882$ & $1.5877\times10^{-4}$ & 1.00000 \\
0.6 & 0.0 & 0 & 0.14 & $0.13838888261$ & $1.3222\times10^{-4}$ & 0.83274 \\
0.6 & 0.3 & 0 & 0.14 & $0.13838325235$ & $1.3337\times10^{-4}$ & 0.84003 \\
0.6 & 0.6 & 0 & 0.14 & $0.13837674451$ & $1.3597\times10^{-4}$ & 0.85640 \\
0.6 & 0.9 & 0 & 0.14 & $0.13837046754$ & $1.4191\times10^{-4}$ & 0.89376 \\
0.0 & 0.0 & 1 & 0.25 & $0.24789116928$ & $7.4363\times10^{-7}$ & 1.00000 \\
0.6 & 0.0 & 1 & 0.25 & $0.24791185305$ & $3.7689\times10^{-7}$ & 0.50683 \\
0.6 & 0.3 & 1 & 0.25 & $0.24791187749$ & $3.3037\times10^{-7}$ & 0.44426 \\
0.6 & 0.6 & 1 & 0.25 & $0.24791190185$ & $2.8091\times10^{-7}$ & 0.37776 \\
0.6 & 0.9 & 1 & 0.25 & $0.24791192477$ & $2.2785\times10^{-7}$ & 0.30640 \\
0.0 & 0.0 & 2 & 0.35 & $0.34749334885$ & $8.6221\times10^{-10}$ & 1.00000 \\
0.6 & 0.0 & 2 & 0.35 & $0.34751031239$ & $2.8126\times10^{-10}$ & 0.32621 \\
0.6 & 0.3 & 2 & 0.35 & $0.34751031351$ & $2.1353\times10^{-10}$ & 0.24765 \\
0.6 & 0.6 & 2 & 0.35 & $0.34751031463$ & $1.4952\times10^{-10}$ & 0.17341 \\
0.6 & 0.9 & 2 & 0.35 & $0.34751031575$ & $8.9346\times10^{-11}$ & 0.10362 \\
\bottomrule
\end{tabular}
\end{table*}

Table~\ref{tab:reference} separates the charge effect from the nonminimal
effect. At \(e=0.6\), changing \(\lambda\) from zero to 0.9
increases the fundamental monopole width at \(\muh=0.14\) by
\(7.33\%\), but reduces the dipole width at \(\muh=0.25\)
to \(60.45\%\) and the quadrupole width at \(\muh=0.35\)
to \(31.77\%\) of their RN values. The respective changes in
real frequency are approximately \(-0.0133\%\),
\(2.89\times10^{-5}\%\), and \(9.68\times10^{-7}\%\).
The lifetime response is therefore not a restatement of a large shift
in binding energy.

Relative to Schwarzschild at the same field mass, the full
\((e,\lambda)=(0.6,0.9)\) widths are \(89.38\%\),
\(30.64\%\), and \(10.36\%\) in these three channels.
Enhanced monopole absorption relative to RN therefore does not imply
enhanced absorption relative to Schwarzschild: charge-induced suppression
remains stronger than the nonminimal enhancement in this comparison.
At \((e,\lambda)=(0.9,0.9)\), the reference quadrupole width is
\(8.5334\times10^{-12}\), about \(0.990\%\) of its
Schwarzschild value, although its real frequency differs by only
\(0.0109\%\). This larger case is not used to infer a
uniform trend arbitrarily close to extremality.

\subsection{Mass-dependent monopole reversal}

\begin{table*}[tb]
\centering
\caption{Multiple field masses at $e=0.6$ and $n_r=0$. The real frequency is for $\lambda=0.9$; the three width columns compare $\lambda=0,0.6,0.9$. $R_{\rm RN}=\Gam(e,\lambda)/\Gam(e,0)$ uses the same charge.}
\label{tab:masses}
\small
\setlength{\tabcolsep}{3pt}
\begin{tabular}{ccrrrrr}
\toprule
$\ell$ & $\muh$ & $\om_R(0.9)$ & $\Gam(0)$ & $\Gam(0.6)$ & $\Gam(0.9)$ & $R_{\rm RN}(0.9)$ \\
\midrule
0 & 0.08 & $0.0797295571$ & $2.6904\times10^{-6}$ & $2.4813\times10^{-6}$ & $2.3899\times10^{-6}$ & 0.88833 \\
0 & 0.12 & $0.1190158810$ & $4.4666\times10^{-5}$ & $4.4448\times10^{-5}$ & $4.5689\times10^{-5}$ & 1.02290 \\
0 & 0.18 & $0.1763604839$ & $7.0898\times10^{-4}$ & $7.5590\times10^{-4}$ & $7.8946\times10^{-4}$ & 1.11351 \\
1 & 0.20 & $0.1989585355$ & $2.0717\times10^{-8}$ & $1.3675\times10^{-8}$ & $9.7173\times10^{-9}$ & 0.46904 \\
1 & 0.30 & $0.2962560012$ & $4.7767\times10^{-6}$ & $3.9530\times10^{-6}$ & $3.5136\times10^{-6}$ & 0.73556 \\
1 & 0.40 & $0.3899055642$ & $3.1904\times10^{-4}$ & $3.0979\times10^{-4}$ & $3.0620\times10^{-4}$ & 0.95976 \\
2 & 0.25 & $0.2491128767$ & $6.4373\times10^{-13}$ & $2.6005\times10^{-13}$ & $1.0886\times10^{-13}$ & 0.16911 \\
2 & 0.35 & $0.3475103158$ & $2.8126\times10^{-10}$ & $1.4952\times10^{-10}$ & $8.9346\times10^{-11}$ & 0.31766 \\
2 & 0.45 & $0.4445296809$ & $4.1966\times10^{-8}$ & $2.7842\times10^{-8}$ & $2.0829\times10^{-8}$ & 0.49633 \\
\bottomrule
\end{tabular}
\end{table*}

\begin{figure*}[tb]
 \centering
 \includegraphics[width=\textwidth]{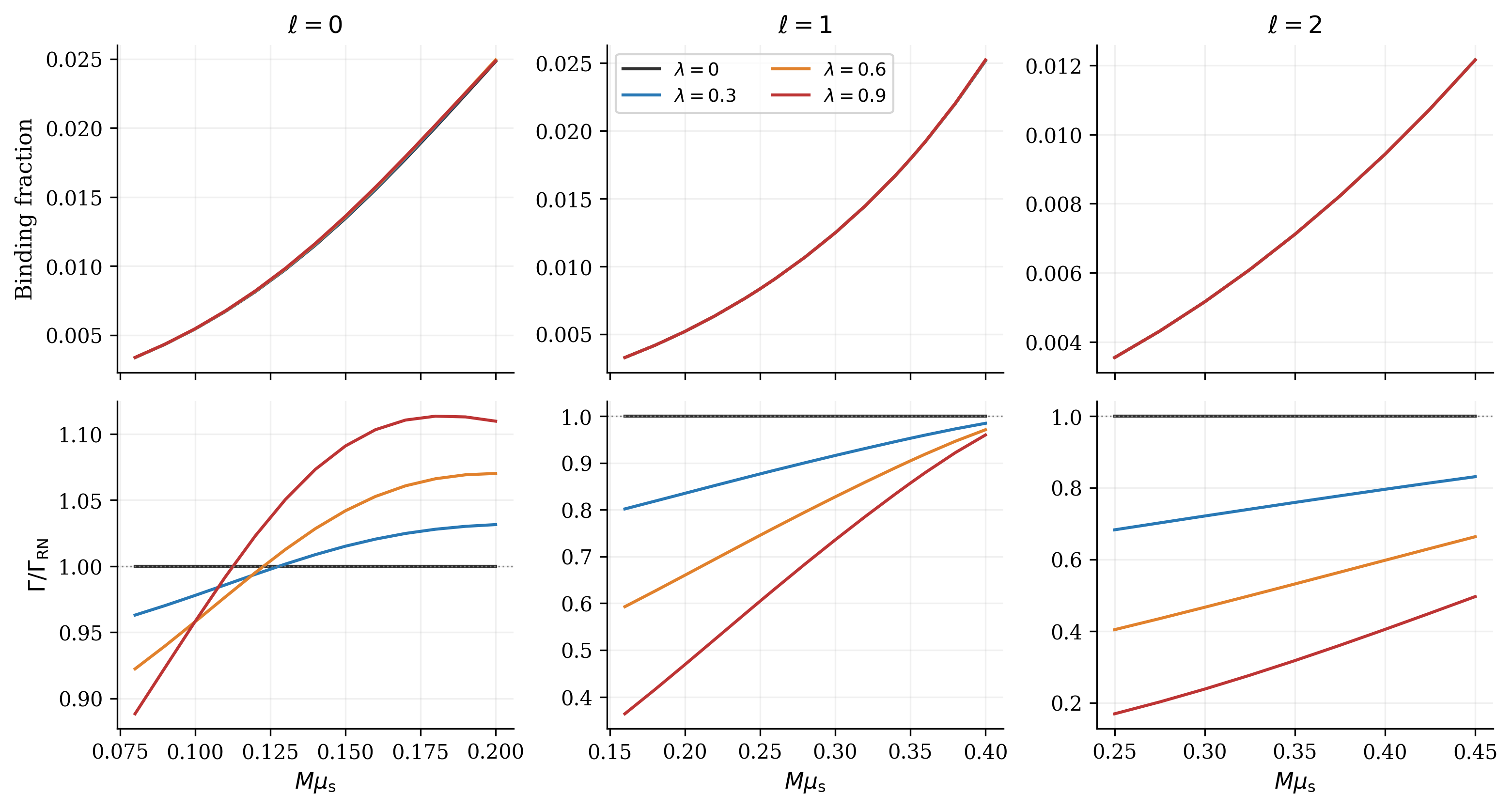}
 \caption{Mass dependence of fundamental clouds at \(e=0.6\).
 Top: binding fractions; the different nonminimal couplings almost overlap
 on this scale. Bottom: amplitude widths divided by the RN width at the
 same field mass and angular number. Only modes passing the trapping and
 precision tests are drawn. The monopole crosses unity, whereas the
 dipole and quadrupole remain suppressed in the displayed intervals.}
 \label{fig:massResponse}
\end{figure*}

The monopole ordering is not uniform in field mass.
Table~\ref{tab:masses} and Fig.~\ref{fig:massResponse} show why a
single reference mass would give an incomplete conclusion. At
\(e=0.6\), \(\lambda=0.9\), the width is about \(88.83\%\)
of the RN width at \(\muh=0.08\), but about \(111.35\%\)
at \(\muh=0.18\). The sign reversal occurs within the retained
trapping interval; it is not caused by losing the exterior well or
switching to a radiative boundary condition.

\begin{table*}[tb]
\centering
\caption{Monopole crossover masses: $\Gam(e,\lambda)=\Gam(e,0)$ for $n_r=0$. The widths are smaller than their RN values below each crossover and larger immediately above it. Digits are deliberately rounded.}
\label{tab:crossovers}
\begin{tabular}{crrr}
\toprule
$e$ & $\lambda=0.3$ & $\lambda=0.6$ & $\lambda=0.9$ \\
\midrule
0.3 & 0.126297 & 0.121304 & 0.111402 \\
0.6 & 0.127917 & 0.122714 & 0.112547 \\
0.9 & 0.121709 & 0.116056 & 0.105843 \\
\bottomrule
\end{tabular}
\end{table*}

We locate the crossing by solving
\begin{equation}
 \Gam_{00}(e,\lambda;\muh_c)
       =\Gam_{00}(e,0;\muh_c)
 \label{eq:crossover}
\end{equation}
along the same continued fundamental branch. Table~\ref{tab:crossovers}
reports nine such comparisons. At \(e=0.6\), the crossing moves
from \(\muh_c=0.127917\) at \(\lambda=0.3\) to
\(0.112547\) at \(\lambda=0.9\). All three sampled charges
exhibit a crossing, and increasing the nonminimal fraction moves it
to lower mass at each fixed charge. We do not extrapolate these nine
roots into a universal monotonic law in charge.

\begin{figure*}[tb]
 \centering
 \includegraphics[width=0.90\textwidth]{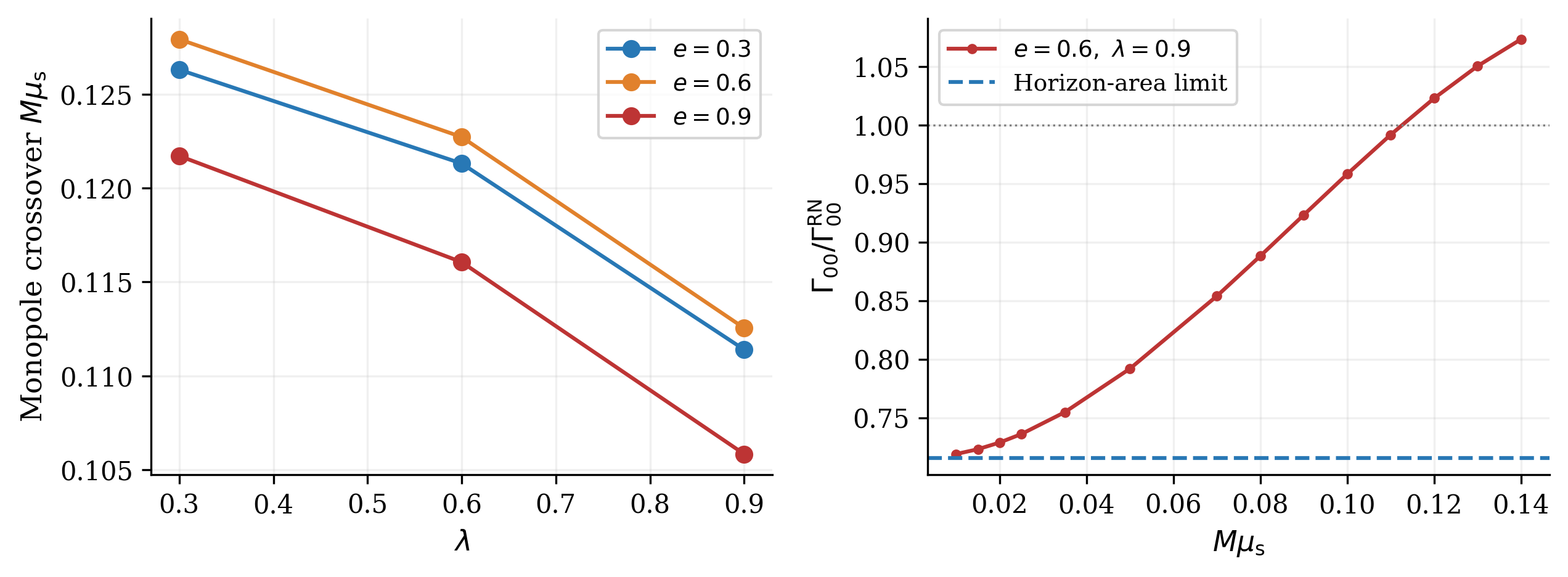}
 \caption{Left: monopole crossover masses at three charges.
 Right: the \(e=0.6\), \(\lambda=0.9\) width ratio approaches the
 horizon-area prediction in the weak-binding limit, then passes through
 unity at finite mass. The dashed line shows the weak-binding asymptote.}
 \label{fig:crossover}
\end{figure*}

The right panel of Fig.~\ref{fig:crossover} connects the reversal with
the small-mass behavior. A smaller horizon reduces the leading low-energy
absorption, but finite-frequency transmission and cloud normalization
eventually overcome that reduction. The dipole and quadrupole examples
instead remain below their RN widths over the plotted mass intervals.
Their suppression also weakens as the field becomes more strongly bound.

\subsection{Charge at fixed nonminimal coupling}

\begin{figure*}[tb]
 \centering
 \includegraphics[width=\textwidth]{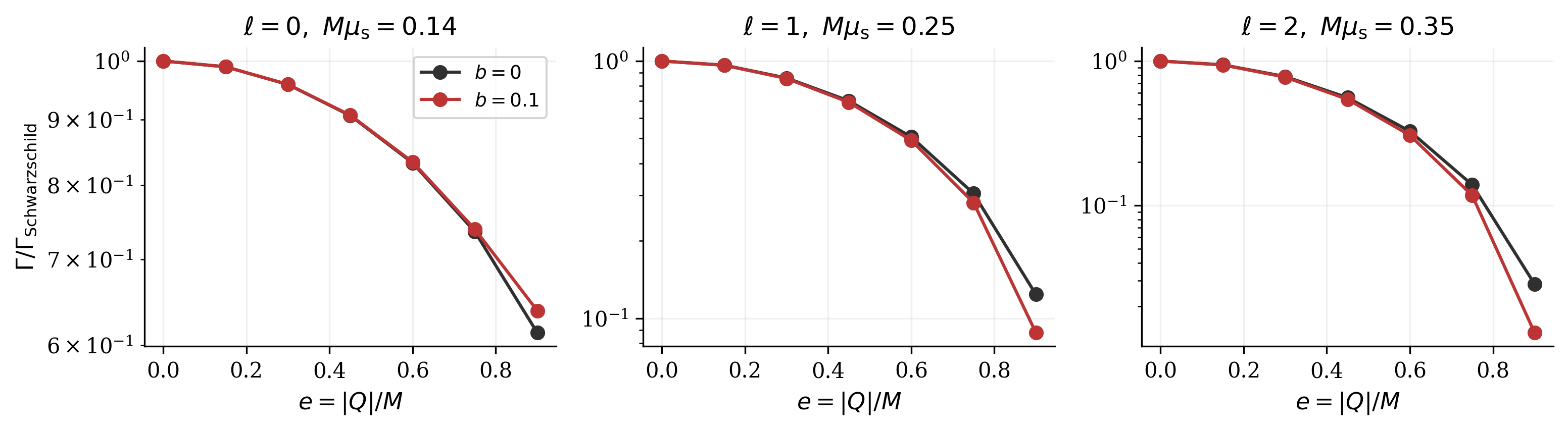}
 \caption{Charge scans at fixed \(b=0\) and \(b=0.1\), with
 \(M\), field mass, and angular number fixed in each panel.
 These 42 additional frequencies separate a change in charge
 from the simultaneous change in \(b\) implicit in a fixed-\(\lambda\)
 scan. The normalization is the Schwarzschild width for each mode.}
 \label{fig:fixedCoupling}
\end{figure*}

Figure~\ref{fig:fixedCoupling} provides a control on the parameterization.
At fixed \(b=0\) or 0.1, increasing \(e\) reduces the widths in
all three sampled reference channels. The effect is particularly strong
for higher angular momentum. Both curves meet Schwarzschild at
\(e=0\), as expected for finite \(b\); keeping \(\lambda>0\)
fixed instead gives the singular coupling limit described in
Sec.~\ref{sec:geometry}. The monopole curves are closely spaced here,
with the sign of the nonminimal correction depending on field mass.

\subsection{Potential wells and radial excitations}

\begin{table*}[tb]
\centering
\caption{Limiting masses $\muh_{\rm well}$ at which the two exterior potential extrema merge. Quasi-bound poles may persist beyond these geometric thresholds.}
\label{tab:wells}
\begin{tabular}{ccrrr}
\toprule
$e$ & $\lambda$ & $\ell=0$ & $\ell=1$ & $\ell=2$ \\
\midrule
0.0 & 0.0 & 0.250000 & 0.465649 & 0.740103 \\
0.6 & 0.0 & 0.258871 & 0.492097 & 0.782486 \\
0.6 & 0.9 & 0.255146 & 0.494016 & 0.785737 \\
0.9 & 0.0 & 0.270644 & 0.538722 & 0.858328 \\
0.9 & 0.9 & 0.268952 & 0.539738 & 0.860173 \\
\bottomrule
\end{tabular}
\end{table*}

The potential-well threshold behaves differently in the monopole and
higher-multipole sectors. At fixed \(e=0.6\), increasing
\(\lambda\) from zero to 0.9 decreases
\(\muh_{\rm well}\) from 0.258871 to 0.255146 for \(\ell=0\),
but increases it from 0.492097 to 0.494016 for \(\ell=1\)
and from 0.782486 to 0.785737 for \(\ell=2\).
Table~\ref{tab:wells} gives the corresponding charge comparisons.
These small changes in well morphology coexist with much larger
relative changes in absorption widths.

The distinction between a well and a trapped level matters numerically.
For example, the Schwarzschild monopole still has a potential well at
\(\muh=0.20<0.25\), but its fundamental pole fails our
three-turning-point condition. Quasi-bound solutions can persist beyond this trapping criterion.
{The neutral survey focuses on the selected trapping spectrum. The charged
mass extension in Sec.~\ref{sec:charged} explicitly also follows poles
outside that selection.}

\begin{figure*}[tb]
 \centering
 \includegraphics[width=0.90\textwidth]{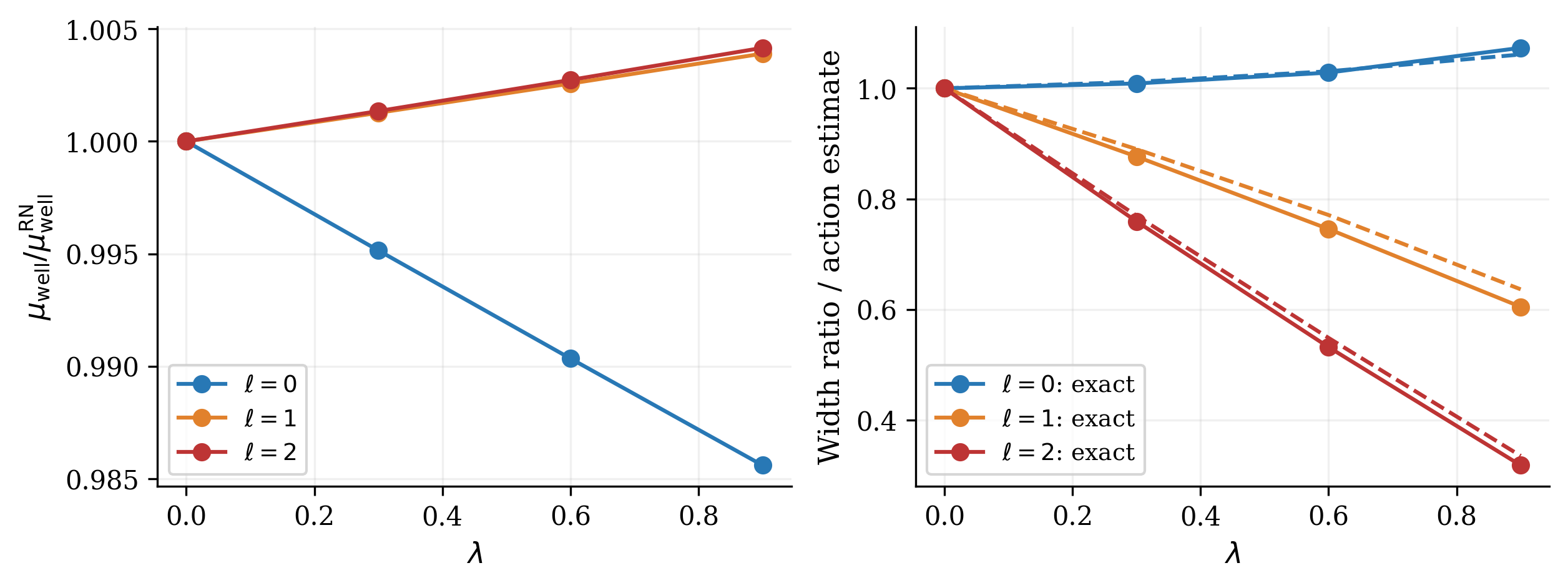}
 \caption{Left: potential-well limiting masses relative to RN at
 \(e=0.6\). Right: exact numerical width ratios (solid curves) and
 the action-only estimates \(\exp[-2(S-S_{\rm RN})]\) (dashed curves)
 at \((\ell,\muh)=(0,0.14),(1,0.25),(2,0.35)\).
 The action captures higher-multipole suppression qualitatively;
 quantitative widths also depend on the matching prefactors.}
 \label{fig:actions}
\end{figure*}

\begin{table*}[tb]
\centering
\caption{First two radial excitations at $e=0.6$. The real frequency is for $\lambda=0.9$; the width ratio compares that geometry with RN at the same $e$, $\muh$, $\ell$, and $n_r$.}
\label{tab:excitations}
\begin{tabular}{cccrrrr}
\toprule
$\ell$ & $\muh$ & $n_r$ & $\om_R(0.9)$ & $\Gam(0)$ & $\Gam(0.9)$ & $R_{\rm RN}$ \\
\midrule
0 & 0.15 & 1 & $0.14952308511$ & $2.6279\times10^{-5}$ & $2.8502\times10^{-5}$ & 1.08462 \\
0 & 0.15 & 2 & $0.14979542384$ & $7.5039\times10^{-6}$ & $8.1156\times10^{-6}$ & 1.08153 \\
1 & 0.25 & 1 & $0.24907887376$ & $1.4106\times10^{-7}$ & $8.5785\times10^{-8}$ & 0.60814 \\
1 & 0.25 & 2 & $0.24948675280$ & $6.3437\times10^{-8}$ & $3.8658\times10^{-8}$ & 0.60940 \\
2 & 0.35 & 1 & $0.34860038291$ & $1.8123\times10^{-10}$ & $5.7936\times10^{-11}$ & 0.31968 \\
2 & 0.35 & 2 & $0.34910761285$ & $1.0922\times10^{-10}$ & $3.5019\times10^{-11}$ & 0.32062 \\
\bottomrule
\end{tabular}
\end{table*}

The first two radial excitations in Table~\ref{tab:excitations} retain
the reference angular ordering. At \(e=0.6\), \(\lambda=0.9\),
the monopole excited-state widths at \(\muh=0.15\) exceed their
RN values, whereas the dipole and quadrupole widths remain smaller.
Higher radial levels lie closer to the rest-mass threshold and decay
more slowly in these examples. Nevertheless their nonminimal/RN width
ratios remain close to those of the fundamental at the same field mass.
This is distinct from an outburst of highly damped QNM overtones:
the excitations here are extended massive bound levels with a different
outer boundary condition.

\begingroup
\section{Charged scalar doublets and triplets}
\label{sec:charged}

\subsection{Gauge representation and angular reduction}

We now extend the gauge-singlet calculation to canonical complex scalar
multiplets in the fundamental and adjoint representations of SU(2).
The background remains the geometry of Eq.~\eqref{eq:metric};
the scalar background is zero. Its stress tensor and gauge current start
at quadratic order, so the linear scalar equation decouples from metric
and Yang--Mills perturbations. No scalar-curvature, Higgs, or additional
nonminimal scalar interaction is introduced.

{``Canonical'' means a positive quadratic kinetic term and a mass
term, with scalar action
\begin{equation}
 S_\Phi=-\int d^4x\sqrt{-g}\,
 \left[(D_\mu\Phi)^\dagger D^\mu\Phi+\mus^2\Phi^\dagger\Phi\right].
 \label{eq:scalarActionReview}
\end{equation}
SU(2) is the special unitary group acting on the internal components.
Its isospin \(t\) labels an internal representation of dimension \(2t+1\): the fundamental doublet has \(t=1/2\)
and the adjoint triplet \(t=1\). A weight \(s\) is an eigenvalue of
\(T_3\). The dagger denotes Hermitian conjugation; \(\nabla_\mu\)
is the spacetime covariant derivative and the gauge coupling is included
in \(\mathscr A_\mu=-ig_{\rm YM}A_\mu^{(a)}T_a\).}

The Wu--Yang connection is effectively Abelian, as identified in
Ref.~\cite{Balakin:2007}. To make the charged model explicit, let
\([T_a,T_b]=i\epsilon_{abc}T_c\), and write the anti-Hermitian
connection as \(\mathscr A\), so \(D_\mu=\nabla_\mu+\mathscr A_\mu\).
For integer winding \(w\), define
\begin{align}
 T_r&=\sin\theta\cos(w\phi)T_1+
       \sin\theta\sin(w\phi)T_2+\cos\theta T_3,\\
 T_\theta&=\partial_\theta T_r,\qquad
 T_\phi=-\sin(w\phi)T_1+\cos(w\phi)T_2.
\end{align}
{Here \(w\) counts the azimuthal winding of the internal radial
direction. The generators \(T_a\) are Hermitian and traceless, and
\(\epsilon_{abc}\) is the fully antisymmetric symbol with
\(\epsilon_{123}=1\).} The connection is
\begin{equation}
 \mathscr A_\theta=iT_\phi,\quad
 \mathscr A_\phi=-iw\sin\theta T_\theta,\quad
 \mathscr A_t=\mathscr A_r=0.
\end{equation}
It satisfies \(D_\theta T_r=D_\phi T_r=0\).
{Here the derivative of a generator is
\(D_\mu T_r=\partial_\mu T_r+[\mathscr A_\mu,T_r]\).
Thus ``reducible'' means that the connection preserves this internal
direction and can be diagonalized locally. We use
\(\Phi'=U^{-1}\Phi\) and
\(\mathscr A'=U^{-1}\mathscr A U+U^{-1}dU\).}
Under \(U=e^{-iw\phi T_3}e^{-i\theta T_2}\), its transformed value is
\(-iw\cos\theta T_3d\phi\). Regular north and south potentials are
\begin{equation}
 \mathscr A_N=iw(1-\cos\theta)T_3d\phi,\qquad
 \mathscr A_S=-iw(1+\cos\theta)T_3d\phi.
\end{equation}
This gauge transformation diagonalizes the background Wu--Yang connection.

For an isospin-\(t\) scalar, the weights \(s=-t,\ldots,t\) see
monopole numbers \(p=ws\). Their patch transition is
\(\Phi_{s,N}=e^{-2ip\phi}\Phi_{s,S}\), with \(2p\in\mathbb Z\).
We set \(w=1\): a fundamental doublet has \(p=\pm1/2\), and a
complex adjoint triplet has \(p=0,\pm1\). The gauge-covariant angular
eigenfunctions are monopole harmonics \cite{WuYang:1976}, with
\begin{equation}
 \Lambda_j=j(j+1)-p^2,\qquad j=|p|,|p|+1,\ldots.
 \label{eq:monopoleHarmonics}
\end{equation}
Their azimuthal labels are \(m=-j,\ldots,j\). All \(m\) values and
both signs of \(p\) have the same radial frequencies in this static
background.

{The north/south patches avoid a gauge singularity on either polar
axis; their single-valued transition enforces \(2p\in\mathbb Z\).
Explicitly, in the diagonal gauge a harmonic has angular dependence
\(e^{im\phi}\Theta(\theta)\), where
\begin{equation}
 -\frac{1}{\sin\theta}\partial_\theta(\sin\theta\,\partial_\theta\Theta)
 +\frac{(m-p\cos\theta)^2}{\sin^2\theta}\Theta=\Lambda_j\Theta.
 \label{eq:angularODEReview}
\end{equation}
Writing \(v=\cos\theta\), set \(a=|m-p|\), \(b_J=|m+p|\), and
\(d_J=j-(a+b_J)/2\). The regular solution is proportional to
\((1-v)^{a/2}(1+v)^{b_J/2}P_{d_J}^{(a,b_J)}(v)\), where
\(P_{d_J}^{(a,b_J)}\) is a Jacobi polynomial and \(d_J\) is a
nonnegative integer. The regular-patch azimuthal factors are
\(e^{i(m-p)\phi}\) and \(e^{i(m+p)\phi}\), respectively. This
construction gives Eq.~\eqref{eq:monopoleHarmonics} and makes the
angular substitution test reproducible.} Half-integer \(j\) labels the combined spatial and gauge angular
momentum of the scalar. Substitution of the Jacobi-polynomial
angular functions checks all 32 states in the first two angular levels
of \(p=0,\pm1/2,\pm1\); the connection identities are checked separately.

The monopole number \(p\) and geometric charge \(e=|Q|/M\) are related
through the background normalization. Restoring Newton's constant and
the gauge coupling in the conventions of
Ref.~\cite{Balakin:2007} gives
\begin{equation}
 Q^2=\frac{4\pi G w^2}{g_{\rm YM}^2},\qquad p=ws.
\end{equation}
At fixed winding and gauge coupling, varying \(e\) varies \(M\).
Holding \(\muh\) and \(b\) fixed simultaneously would then vary the
particle mass and dimensional nonminimal coupling. Cross-charge comparisons
in this section therefore compare dimensionless parameter sets. The lifetime-response comparisons
hold \(e,M,\muh\) fixed and vary \(\xi\).

\subsection{Radial problem and limiting behavior}

For \((D_\mu D^\mu-\mus^2)\Phi=0\), separation gives
\begin{equation}
 \frac{d^2u}{dr_*^2}+
 \left\{\omega^2-f\left[\mus^2+\frac{\Lambda_j}{r^2}
                              +\frac{f'}r\right]\right\}u=0.
 \label{eq:chargedRadial}
\end{equation}
The ingoing and decreasing conditions remain Eq.~\eqref{eq:boundaries},
with no electric \(\omega-qQ/r\) shift. Introduce the effective radial
order \(L=(\sqrt{1+4\Lambda_j}-1)/2\), so that
\(L(L+1)=\Lambda_j\). The auxiliary order \(L\) is generally noninteger, while \(j\)
retains its role as the physical angular quantum number. Both radial solvers use the exact
angular eigenvalue, without rounding \(L\). The weak-binding limit is
\begin{equation}
 \frac{\om_R}{\muh}=1-\frac{\muh^2}{2\mathcal N^2}+O(\muh^4),
 \qquad \mathcal N=n_r+L+1.
 \label{eq:chargedHydrogenic}
\end{equation}
The definitions of width, binding, and lifetime in
Eq.~\eqref{eq:observables} are unchanged. We apply the same
three-turning-point criterion, Eq.~\eqref{eq:trapping}, with the charged
potential. Since \(\Lambda_j\geq0\), the positive-potential proof
in Sec.~\ref{sec:field} also excludes growing modes of these canonical
scalar sectors. This does not establish stability of coupled EYM
perturbations. {Superradiance is reflected-wave amplification through
extraction of rotational or electric energy from the background.}
Rotating magnetic-black-hole superradiance
\cite{Pereniguez:2024} cannot be transferred to this static background.

An independent weak-binding check follows from the horizon-normalized
static scalar amplitude \(\varphi=u/x\):
\begin{equation}
 (x^2f\varphi')'=\Lambda_j\varphi,\qquad
 \varphi(x_+)=1,\quad
 \varphi'(x_+)=\frac{\Lambda_j}{x_+^2f_x(x_+)}.
\end{equation}
Define \(C_L\) by \(\varphi\sim C_Lx^L\) at large radius. Matching
to the hydrogenic \(n_r=0\) wave and dividing the horizon flux by twice
the exterior norm yields
\begin{align}
 \Gam&\sim K_L\muh^{4L+6},\\
 K_L&=\frac{x_+^2}{2C_L^2\Gamma(2L+3)}
                  \left(\frac2{L+1}\right)^{2L+3}.
 \label{eq:chargedWidthLimit}
\end{align}
Here \(\Gamma(\cdot)\) is Euler's gamma function. For  RN,
\(d=\sqrt{1-e^2}\), the static solution is
\(P_L((x-1)/d)\), giving
\begin{equation}
 C_L^{\rm RN}=\frac{\Gamma(2L+1)}{2^L\Gamma(L+1)^2d^L}.
\end{equation}
At \(L=0\), \(C_0=1\), recovering Eq.~\eqref{eq:areaLimit}.
The lowest charged doublet and triplet instead have mass powers
\(4L+6=7.4641016\) and \(8.4721360\). Their leading absorption depends on both the transfer coefficients
and the horizon area.

{The transfer coefficient \(C_L\) measures how a horizon-normalized
static solution grows toward the far region. In an overlap region,
write the ground-state outer solution as
\(u\simeq A x^{L+1}e^{-\kappa_B x}\), with
\(\kappa_B=\muh^2/(L+1)\). The norm is dominated by the far cloud, where \(f\simeq1\); integrating
\(|A|^2x^{2L+2}e^{-2\kappa_B x}\) gives
\(I\simeq |A|^2\Gamma(2L+3)/(2\kappa_B)^{2L+3}\), and matching
\(\varphi\sim C_Lx^L\) gives \(u_H\simeq A x_+/C_L\).
Horizon flux balance in the weak-leakage approximation then yields
\(\Gam\simeq |u_H|^2/(2I)\), giving Eq.~\eqref{eq:chargedWidthLimit}.
This matching relation applies for \(n_r=0\) as \(\muh\to0\)
at fixed nonextremal geometry.
Here \(P_L\) in the RN solution is the Legendre function of the first
kind, distinct from the Jacobi polynomial above.

For the numerical static transfer, the horizon expansion starts at
\(x_++10^{-6}\), using the displayed value and slope. DOP853 integration
with tolerances \(2\times10^{-12}\) and \(2\times10^{-14}\) reaches
\(X=10^4,2\times10^4,4\times10^4,8\times10^4,1.6\times10^5\).
We extrapolate
\(C_L(X)=[X\varphi'(X)+(L+1)\varphi(X)]/[(2L+1)X^L]\)
by a cubic fit in \(1/X\). This estimator cancels the leading
\(X^{-L-1}\) static solution. Doubling the horizon offset tests the
initialization, while the analytic RN coefficient tests the far-radius
extraction.}

\subsection{Accuracy and charged-channel lifetime reversals}

The initial charged survey contains 304 charged-sector representative
poles and 18 neutral-weight controls. It is separate from the neutral
survey counts in Sec.~\ref{sec:results}; controls and the two signs of
\(p\) are not counted as new independent charged modes. The survey covers
\(e=0.3,0.6,0.9\), with a finer lowest-channel scan
\(0.08\leq\muh\leq0.32\) at \(e=0.6\), and selected higher angular
states through \(\muh=0.35\). Not all parameter combinations are sampled.

{The initial charged integration uses the neutral tolerances and
boundary orders, but places the match at
\(\max[10,\mathcal N^2/(2\muh^2)]\), with
\(\max[12,0.65\mathcal N^2/\muh^2]\) in the boundary check.
RN seeds are found with a depth-600 fraction, then continued in six
steps of nonminimal fraction. Subsequent mass steps are at most 0.01;
the real frequency is predicted to scale with mass and the width with
\(\muh^{4L+6}\), before solving the full equation. These predictors
and the principal-number diagnostic guide branch selection before the
numerical eigenvalue condition is imposed.}

Every pole in this initial survey is recomputed with changed horizon offset, outer radius,
matching radius, series order, and integration tolerance. The maximum
relative width change is \(2.9\times10^{-7}\), and the maximum absolute
frequency change is \(2.0\times10^{-13}\). All 322 poles pass the
adopted empirical thresholds of \(10^{-6}\) and \(10^{-10}\),
respectively; 303 also pass the cloud criterion. The other 19 remain
in the data but are not used as three-turning-point clouds.

Twenty charged poles, including higher angular states and radial
excitations, are independently checked with 60-digit continued fractions
at depths 600, 1200, 2400, and 4800. The largest CF--ODE relative width
difference is \(2.2\times10^{-10}\); the largest change between the
last two depths is \(3.7\times10^{-8}\).
Table~\ref{charged:tab:validation} gives the strongest-deformation checks.
Twelve overlapping neutral-weight controls recover the saved neutral
spectra to \(3.1\times10^{-11}\) in relative width. These comparisons
provide empirical evidence of convergence.

\begin{table*}[tb]
\centering
\caption{Independent checks at $\lambda=0.9$. The two errors are relative differences in the amplitude width: CF at depth 4800 versus the primary ODE, and CF depth 4800 versus 2400. All twenty checked cases and all four depths are retained in the data.}
\label{charged:tab:validation}
\begin{tabular}{ccccccc}
\toprule
$e$ & $|p|$ & $j$ & $n_r$ & $\muh$ & CF/ODE difference & CF depth change \\
\midrule
0.6 & 0.5 & 0.5 & 0 & 0.14 & $1.6\times10^{-11}$ & $7.2\times10^{-9}$ \\
0.6 & 0.5 & 0.5 & 0 & 0.22 & $1.9\times10^{-11}$ & $3.0\times10^{-14}$ \\
0.6 & 1 & 1 & 0 & 0.14 & $1.5\times10^{-11}$ & $2.1\times10^{-8}$ \\
0.6 & 1 & 1 & 0 & 0.22 & $1.3\times10^{-11}$ & $1.1\times10^{-13}$ \\
0.9 & 0.5 & 0.5 & 0 & 0.22 & $1.8\times10^{-11}$ & $1.6\times10^{-11}$ \\
0.9 & 1 & 1 & 0 & 0.22 & $9.5\times10^{-12}$ & $9.3\times10^{-12}$ \\
0.6 & 0.5 & 1.5 & 0 & 0.30 & $3.6\times10^{-11}$ & $9.7\times10^{-14}$ \\
0.6 & 1 & 2 & 0 & 0.30 & $2.2\times10^{-10}$ & $1.0\times10^{-12}$ \\
0.6 & 0.5 & 0.5 & 1 & 0.20 & $2.3\times10^{-11}$ & $9.7\times10^{-10}$ \\
0.6 & 0.5 & 0.5 & 2 & 0.20 & $4.2\times10^{-11}$ & $3.7\times10^{-8}$ \\
0.6 & 1 & 1 & 1 & 0.20 & $1.6\times10^{-11}$ & $1.3\times10^{-10}$ \\
0.6 & 1 & 1 & 2 & 0.20 & $1.8\times10^{-11}$ & $2.5\times10^{-9}$ \\
\bottomrule
\end{tabular}
\end{table*}

\begin{table*}[tb]
\centering
\caption{Lowest charged channels at $e=0.6$, $j=|p|$, and $n_r=0$. The last column compares the amplitude width with  RN at the same mass and angular sector. All listed poles satisfy the cloud criterion.}
\label{charged:tab:reference}
\begin{tabular}{cccccc}
\toprule
$|p|$ & $\muh$ & $\lambda$ & $\om_R$ & $\Gam$ & $\Gam/\Gam_{\rm RN}$ \\
\midrule
0.5 & 0.14 & 0 & 0.13921702398 & $1.8948\times10^{-6}$ & 1.000000 \\
0.5 & 0.14 & 0.6 & 0.13921694974 & $1.5819\times10^{-6}$ & 0.834877 \\
0.5 & 0.14 & 0.9 & 0.13921685978 & $1.3979\times10^{-6}$ & 0.737760 \\
0.5 & 0.22 & 0 & 0.21656342212 & $1.4405\times10^{-4}$ & 1.000000 \\
0.5 & 0.22 & 0.6 & 0.21655097707 & $1.4171\times10^{-4}$ & 0.983756 \\
0.5 & 0.22 & 0.9 & 0.21654381601 & $1.4248\times10^{-4}$ & 0.989046 \\
1 & 0.14 & 0 & 0.13945694652 & $6.5454\times10^{-8}$ & 1.000000 \\
1 & 0.14 & 0.6 & 0.13945695481 & $4.6856\times10^{-8}$ & 0.715855 \\
1 & 0.14 & 0.9 & 0.13945695835 & $3.5214\times10^{-8}$ & 0.537993 \\
1 & 0.22 & 0 & 0.21774999987 & $8.3463\times10^{-6}$ & 1.000000 \\
1 & 0.22 & 0.6 & 0.21774959037 & $7.1983\times10^{-6}$ & 0.862450 \\
1 & 0.22 & 0.9 & 0.21774925276 & $6.6022\times10^{-6}$ & 0.791033 \\
\bottomrule
\end{tabular}
\end{table*}

Table~\ref{charged:tab:reference} illustrates the lowest charged channels.
At \(e=0.6\), \(\lambda=0.9\), and \(\muh=0.20\), their width
ratios to  RN are approximately \(0.936\) and \(0.730\) for
the doublet and charged triplet. The response is mass dependent:
Fig.~\ref{fig:chargedResponse} shows suppression crossing to enhancement
and then back to suppression when the same poles are continued to larger
field mass. Fractional binding changes much less strongly in the original
trapped-cloud interval, but this comparison need not persist on the broader
high-mass resonances.

\begin{figure*}[tbp]
 \centering
 \includegraphics[width=\textwidth]{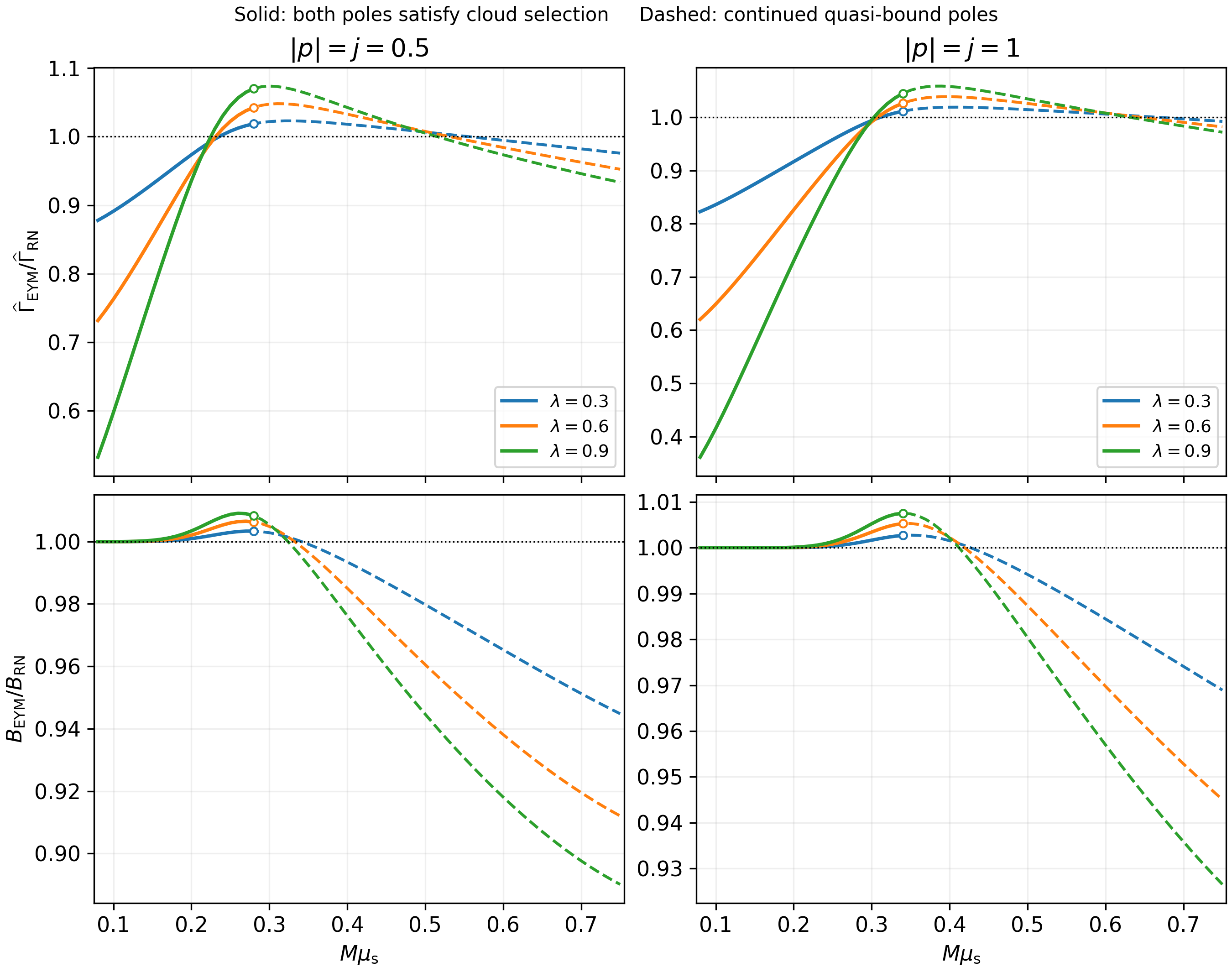}
 \caption{Nonminimal response of the lowest charged channels at
 \(e=0.6\), \(n_r=0\): amplitude-width ratios (top) and
 fractional-binding ratios (bottom), relative to RN, extended
 through \(\muh=0.75\). Solid curves require both poles to satisfy the
 three-turning-point cloud criterion; dashed curves continue the same
 ingoing/decaying quasi-bound branches when at least one pole fails that
 criterion. Open circles mark the last sampled pair passing the cloud selection.
 All plotted poles pass the numerical accuracy checks; the high-mass
 dashed portions describe broader resonances beyond the trapping region.}
 \label{fig:chargedResponse}
\end{figure*}

To determine what happens beyond the original curve endings, we continue
all eight lowest-channel branches at \(e=0.6\), including RN, from
\(\muh=0.32\) through \(0.75\). This adds 344 poles on a
\(\Delta\muh=0.01\) output grid, with smaller continuation steps where
needed and no rounding of the effective angular order. Branch tracking
uses the preceding complex frequencies throughout the strong-binding region. Every added pole is checked
with altered boundary data; larger horizon offsets and higher series
orders control the loss of precision for broader resonances.

{For this extension, through \(\muh=0.60\), the primary/check settings
are horizon offsets \(10^{-3}/2\times10^{-3}\), orders \(12/14\),
outer multipliers \(35/45\), and relative tolerances
\(3\times10^{-13}/3\times10^{-14}\). The matching radii are
\(\max[5,(L+1)^2/(2\muh^2)]\) and
\(\max[6,0.65(L+1)^2/\muh^2]\). Above 0.60 the offsets are
\(0.01/0.02\) and orders \(14/16\). The broadest doublet endpoint
at \(\lambda=0.9\), \(\muh=0.75\) is refined with offsets
\(0.03/0.04\), orders \(18/20\), and tolerance \(3\times10^{-14}\)
on both sides of the comparison. Mass continuation uses a secant
prediction from the preceding two poles, steps no larger than 0.005
below 0.60 and 0.01 above, and step halving on a failed solve or excessive
predictor correction. Each endpoint is checked by a single reverse-continuation step.} Only eight
of the added poles pass the cloud selection. All nevertheless retain
\(\om_I<0\), \(0<\om_R<\muh\), and the decreasing-infinity sheet
\(\Re k>0\). The continuation is checked independently with 60-digit
continued fractions at \(\muh=0.40,0.50,0.60,0.75\) on each branch,
using depths 600, 1200, and 2400. Across the added poles, boundary
variation changes the complex frequency by at most \(9.7\times10^{-11}\)
and the width by \(1.2\times10^{-9}\) relatively. For the 32 independently
checked cases, CF--ODE differences are below \(5.8\times10^{-11}\) in
frequency and \(2.7\times10^{-10}\) in relative width; the last CF depth
change in width is below \(6.4\times10^{-18}\). These comparisons quantify the empirical convergence of the extended survey.

The larger mass interval reveals a turnover followed by a second
width crossing. For \(\lambda=0.3,0.6,0.9\), respectively, the sampled
width-ratio maxima occur near \(\muh=(0.32,0.31,0.30)\) in the
lowest doublet and \((0.41,0.40,0.39)\) in the lowest charged triplet.
The peak locations are estimated from the output grid. Direct equal-width
root finding locates the subsequent return to suppression at
\(\muh\simeq(0.5560,0.5317,0.5152)\) for the doublet and
\((0.6661,0.6449,0.6306)\) for the triplet. Altered-boundary checks
leave an equal-width ratio residual below \(2.5\times10^{-10}\).
All six second crossings fail the cloud criterion in both geometries:
they occur on the broader quasi-bound branches beyond the trapped-cloud region. At \(\muh=0.75\) and
\(\lambda=0.9\), the width ratios are approximately \(0.9334\) and
\(0.9719\), whereas the binding ratios are \(0.8901\) and \(0.9266\).
The absolute widths increase along every sampled branch in this extended
interval. The falling EYM/RN ratio therefore describes relative suppression
along branches whose absolute lifetimes decrease with mass. The tunneling
interpretation remains restricted to narrow, trapped resonances.

In the original cloud-selected survey, twelve crossings are located by
solving the equal-width condition directly
and checking both roots with altered boundaries
(Table~\ref{charged:tab:crossovers}). At \(e=0.6\), \(\lambda=0.9\),
the crossing masses are \(\muh\simeq0.224729\) and \(0.302953\)
for \(|p|=1/2\) and \(|p|=1\). All twelve crossings are trapped
in both geometries; the largest refined width-ratio residual is
\(4.6\times10^{-11}\).
{For these charged crossings we bracket a sign change
of \(\log(\Gam_{\rm EYM}/\Gam_{\rm RN})\) between sampled masses,
seed both complex roots by interpolation on their own branches, and
apply Brent's bracketed method with mass tolerance \(2\times10^{-10}\).
The neutral crossings use the ratio minus unity, bracketed on
\([0.06,0.18]\), with tolerance \(2\times10^{-9}\); the second,
high-mass charged crossings use the same latter tolerance. Each equality
is rechecked by solving both poles with the altered boundaries.} For \(e=0.9\), no triplet crossing is
bracketed before the sampled endpoint \(\muh=0.30\); no conclusion
about an uncomputed crossing beyond that endpoint is drawn.

\begin{table*}[tb]
\centering
\caption{Masses where $\Gam_{\rm EYM}=\Gam_{\rm RN}$ in the lowest charged channels. Every located crossing satisfies the three-turning-point criterion in both geometries. \emph{Not found} means no sign change in the sampled interval $0.10\leq\muh\leq0.30$ at $e=0.9$; crossings outside this interval remain untested.}
\label{charged:tab:crossovers}
\begin{tabular}{cccc}
\toprule
$e$ & $\lambda$ & $|p|=1/2$ & $|p|=1$ \\
\midrule
0.3 & 0.6 & 0.223965 & 0.296882 \\
0.3 & 0.9 & 0.217608 & 0.292107 \\
0.6 & 0.3 & 0.235461 & 0.311430 \\
0.6 & 0.6 & 0.230987 & 0.307564 \\
0.6 & 0.9 & 0.224729 & 0.302953 \\
0.9 & 0.6 & 0.239171 & not found \\
0.9 & 0.9 & 0.234334 & not found \\
\bottomrule
\end{tabular}
\end{table*}

\subsection{Comparison at fixed total angular momentum}

A complementary comparison holds \(j\) fixed across the triplet sectors. The lowest charged channel has
\(\Lambda_j=|p|\), so it has a barrier absent from a neutral monopole.
Within the triplet at \(j=1\), however, the neutral weight has
\(\Lambda_j=2\), while the charged weights have \(\Lambda_j=1\).
Magnetic coupling then reduces the angular barrier.

Table~\ref{charged:tab:samej} and Fig.~\ref{fig:chargedSameJ} quantify
the resulting enhancement of absorption. At \(e=0.6\),
\(\lambda=0.9\), the charged-to-neutral width ratios are about
385, 181, and 87 for \(\muh=0.14,0.22,0.30\), respectively.
The corresponding fractional-binding ratios are about 1.55, 1.61,
and 1.73. Thus a gauge charge does not universally lengthen a cloud's
lifetime. There is no neutral half-integer scalar harmonic with which
to make a same-\(j=1/2\) comparison for the lowest doublet.

\begin{table*}[tb]
\centering
\caption{Within the adjoint triplet at fixed $j=1$, $e=0.6$, $\lambda=0.9$, and $n_r=0$: charged and neutral weights have $\Lambda=1$ and $2$, respectively. Every listed pole is trapped. Subscripts $0$ and $1$ denote $|p|$.}
\label{charged:tab:samej}
\begin{tabular}{ccccc}
\toprule
$\muh$ & $\Gam_0$ & $\Gam_1$ & $\Gam_1/\Gam_0$ & $B_1/B_0$ \\
\midrule
0.10 & $1.572\times10^{-12}$ & $1.043\times10^{-9}$ & 663.47 & 1.5400 \\
0.14 & $9.143\times10^{-11}$ & $3.521\times10^{-8}$ & 385.16 & 1.5532 \\
0.18 & $2.342\times10^{-9}$ & $5.962\times10^{-7}$ & 254.52 & 1.5744 \\
0.22 & $3.648\times10^{-8}$ & $6.602\times10^{-6}$ & 180.99 & 1.6085 \\
0.26 & $4.054\times10^{-7}$ & $5.310\times10^{-5}$ & 130.99 & 1.6628 \\
0.30 & $3.514\times10^{-6}$ & $3.067\times10^{-4}$ & 87.28 & 1.7345 \\
\bottomrule
\end{tabular}
\end{table*}

\begin{figure*}[tb]
 \centering
 \includegraphics[width=\textwidth]{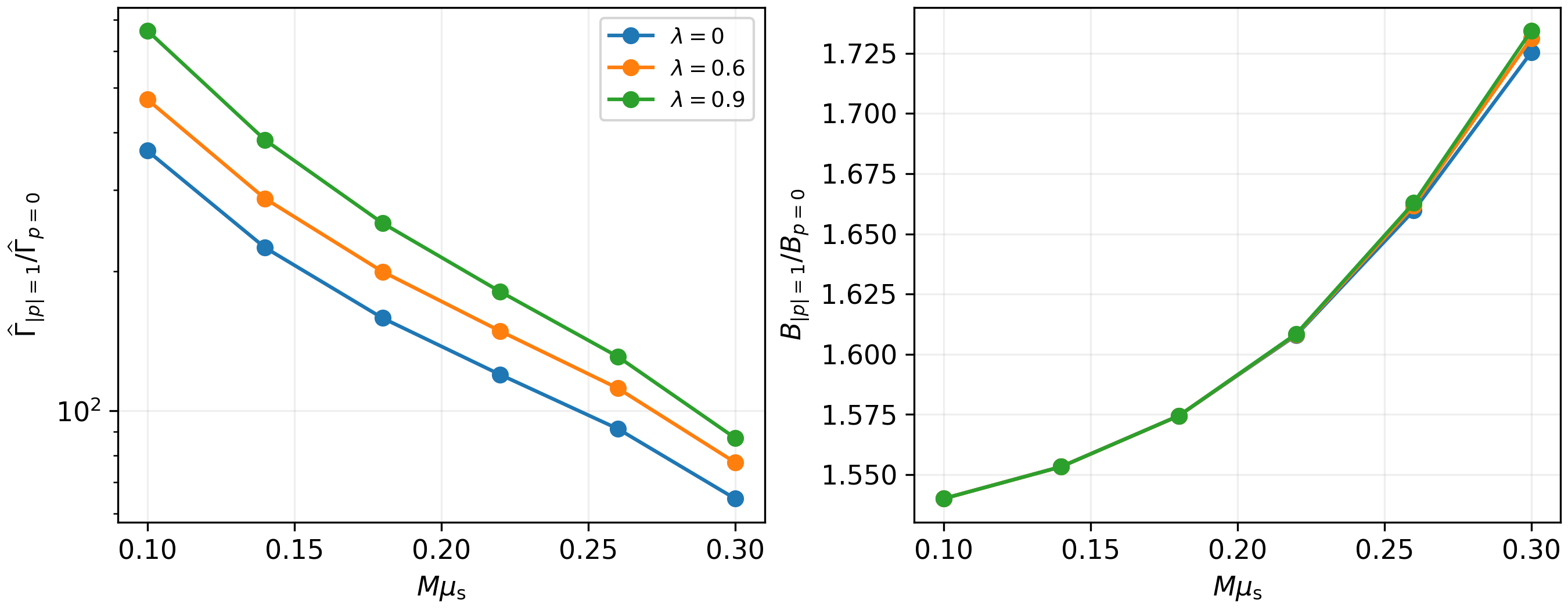}
 \caption{Charged-to-neutral ratios within the adjoint triplet at fixed
 \(j=1\), \(n_r=0\), and \(e=0.6\). The charged angular eigenvalue
 is smaller, and both binding and absorption increase.}
 \label{fig:chargedSameJ}
\end{figure*}

\subsection{Weak binding and radial excitations}

The static-transfer integration recovers the analytic RN coefficient to
better than \(3\times10^{-12}\) in relative value at \(e=0.6\).
{Doubling the static horizon offset from \(10^{-6}\)
to \(2\times10^{-6}\) changes the coefficient by less than
\(2\times10^{-10}\).} Table~\ref{charged:tab:static} gives the
coefficients for both lowest charged channels. At \(\lambda=0.9\),
the width-prefactor ratios to RN are approximately 0.3895 and 0.2510:
one geometric horizon-area ratio cannot describe both channels.

Twenty supplemental weak-binding calculations cover
\(\muh=0.03,\allowbreak 0.04,\allowbreak 0.05,\allowbreak 0.06,\allowbreak 0.08\), with maximum relative boundary
width change \(5.1\times10^{-7}\). Figure~\ref{fig:chargedWeak}
shows approach to both leading limits. At \(\muh=0.03\), the widths
still exceed the leading asymptotic prediction by about 6--13\%; these
finite-mass corrections are much larger than the numerical error.

\begin{table*}[tb]
\centering
\caption{Static transfer coefficients and leading width prefactors at $e=0.6$, $j=|p|$, $n_r=0$. The final column is the actual spectral width divided by Eq.~\eqref{eq:chargedWidthLimit} at $\muh=0.03$, quantifying the finite-mass corrections.}
\label{charged:tab:static}
\begin{tabular}{ccccc}
\toprule
$|p|$ & $\lambda$ & $C_L$ & $K_L$ & ratio \\
\midrule
0.5 & 0 & 0.97346387 & 1.63785895 & 1.06802 \\
1 & 0 & 1.04794429 & 0.44508531 & 1.05993 \\
0.5 & 0.9 & 1.31993528 & 0.63789466 & 1.13002 \\
1 & 0.9 & 1.76981942 & 0.11173772 & 1.13065 \\
\bottomrule
\end{tabular}
\end{table*}

\begin{figure*}[tb]
 \centering
 \includegraphics[width=\textwidth]{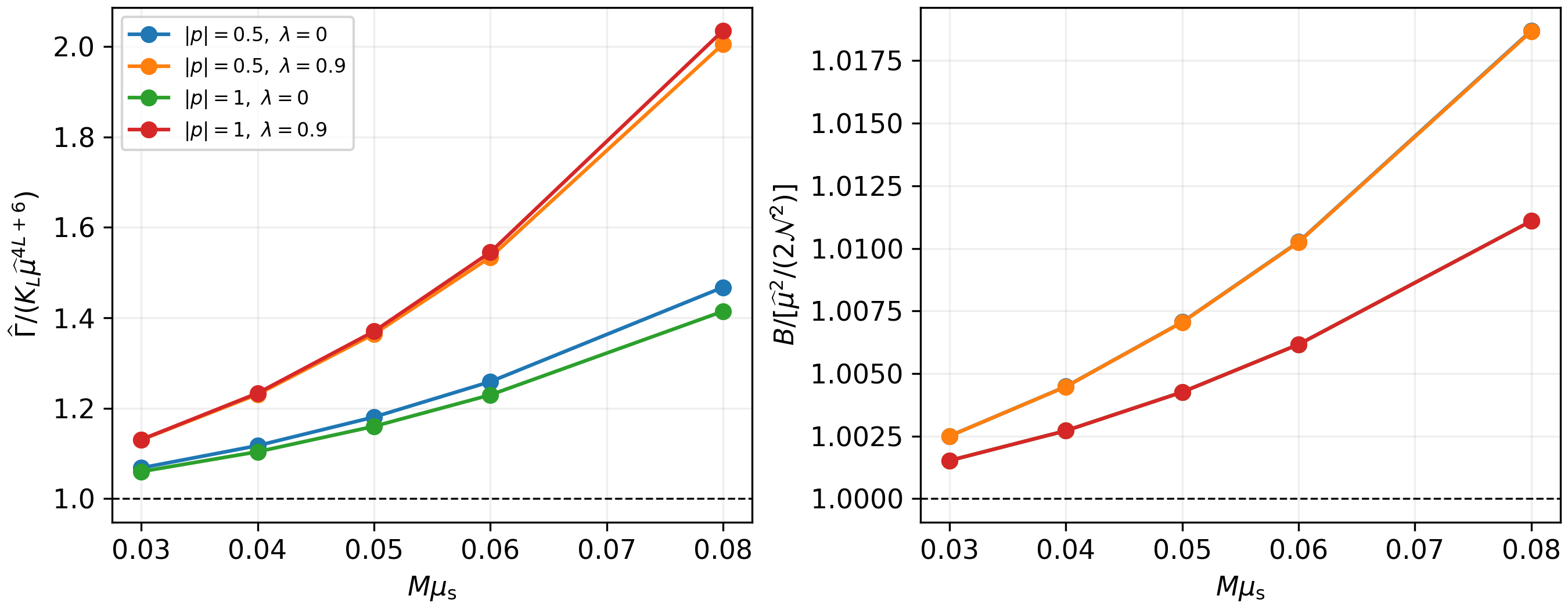}
 \caption{Approach to the independent weak-binding width and energy
 limits at \(e=0.6\), \(j=|p|\), \(n_r=0\). The horizontal line shows
 the weak-binding asymptote.}
 \label{fig:chargedWeak}
\end{figure*}

Finally, Table~\ref{charged:tab:radial} gives the first two radial
excitations at \(\muh=0.20\). Binding and widths decrease with
excitation, while their nonminimal-to-RN width ratios stay close to those
of the respective fundamental states. Independent continued fractions
check all four deformed excited poles. These excitations follow the hydrogenic radial sequence.

\begin{table*}[tb]
\centering
\caption{Radial excitations at $e=0.6$, $\lambda=0.9$, $\muh=0.20$, and $j=|p|$. The radial label $n_r$ follows the hydrogenic sequence. All listed poles are trapped.}
\label{charged:tab:radial}
\begin{tabular}{ccccc}
\toprule
$|p|$ & $n_r$ & $\om_R$ & $\Gam$ & $\Gam/\Gam_{\rm RN}$ \\
\midrule
0.5 & 0 & 0.1975051218 & $5.313\times10^{-5}$ & 0.93568 \\
0.5 & 1 & 0.1992003798 & $1.126\times10^{-5}$ & 0.94080 \\
0.5 & 2 & 0.1996154816 & $3.893\times10^{-6}$ & 0.94147 \\
1 & 0 & 0.1983447482 & $2.073\times10^{-6}$ & 0.72992 \\
1 & 1 & 0.1993771365 & $5.677\times10^{-7}$ & 0.73371 \\
1 & 2 & 0.1996781435 & $2.211\times10^{-7}$ & 0.73480 \\
\bottomrule
\end{tabular}
\end{table*}

\endgroup

\section{Interpretation and scope}
\label{sec:discussion}

\subsection{Why the monopole response can reverse}

The asymptotic expansion \eqref{eq:farMetric} explains why the level
energies are comparatively insensitive to the nonminimal coupling. The
leading \(1/x\) attraction is unchanged, and the first explicit
nonminimal correction is far down the inverse-radius expansion. The
imaginary part, however, depends on matching an extended cloud to a
near-horizon absorbing solution. An apparently small change in a barrier
can therefore dominate a fractional change in the width.

The weak-binding monopole limit supplies a useful analytic anchor.
At zero frequency and zero field mass, the regular \(\ell=0\) solution
has \(u\propto r\), or constant \(\Phi\). This is the same low-energy
behavior underlying the horizon-area absorption law
\cite{Das:1997}. In the overlap region with the normalized hydrogenic
ground state, write \(u\simeq A r\exp(-r/a_0)\), where
\(a_0=(M\mus^2)^{-1}\). Its outer norm is
\(\int |u|^2dr\simeq |A|^2a_0^3/4\), while the ingoing horizon
amplitude is \(u_H\simeq A r_+\). Equating the horizon loss to the
slow change of the cloud norm gives, to leading order at fixed
nonextremal geometry,
\begin{equation}
 \Gam_{00}\sim 2x_+^2\muh^6,\qquad
 \frac{\Gam_{00}(e,b)}{\Gam_{00}(e,0)}
       \longrightarrow\left[\frac{x_+(e,b)}{x_+(e,0)}\right]^2.
 \label{eq:areaLimit}
\end{equation}
The factor of two is for the amplitude-width convention
\eqref{eq:observables}. This quasistationary matching argument uses the norm of the exterior
cloud in the weak-leakage limit. The limits \(\muh\to0\) and extremality need not commute.

Since the horizon shrinks when \(b\) increases, Eq.~\eqref{eq:areaLimit}
predicts suppression of monopole absorption at sufficiently small field
mass. At finite \(\muh\), changes in the inner wave profile and cloud
normalization modify this ordering. The measured crossover marks the
point at which finite-frequency transmission overcomes the leading area effect. For higher angular momenta, the
centrifugal barrier adds a further suppression mechanism.

For three-turning-point states we also compute the dimensionless
forbidden-region action
\begin{equation}
 S=\int_{x_1}^{x_2}\frac{\sqrt{\mathcal V(x)-\om_R^2}}{f(x)}\,dx,
 \qquad \mathcal V(x_i)=\om_R^2.
 \label{eq:action}
\end{equation}
The semiclassical factor \(e^{-2S}\) indicates the strength of barrier
penetration and provides a qualitative interpretation of the numerical
spectra. Accurate absolute widths, especially for monopoles with small
action, also require the oscillation period and matching prefactors.

{In Fig.~\ref{fig:actions}, each action is evaluated at that geometry's
own real eigenfrequency. The action-only ratio
\(\exp[-2(S-S_{\rm RN})]\) deliberately omits the oscillation-time
and amplitude-matching prefactors. We integrate between bracketed turning
points with adaptive quadrature and absolute tolerance \(10^{-10}\).
}

\subsection{Relation to radiative probes}

The QNM overtone sensitivity reported in Ref.~\cite{Lutfuoglu:2025ljm}
and the lifetime sensitivity found here both emphasize the role of the
inner geometry, but they are different spectral statements. QNMs continue
an outgoing radiation condition; our clouds lie on the decreasing branch
\eqref{eq:boundaries}. The radial excitations in our tables remain
connected to the weak-binding hydrogenic levels.
Similarly, a grey-body factor is a real-frequency scattering quantity
above the propagation threshold, whereas a quasi-bound width is the
imaginary part of a pole below that threshold. A monotonic increase in a
grey-body factor at chosen radiative frequencies does not fix the sign of
a below-threshold width change. Comparisons also require the mass/horizon
unit conversion described in Sec.~\ref{sec:geometry}.

The massive-field work of Ref.~\cite{Dubinsky:2025EYM} provides especially
relevant context. Here the square-root sheet, radial labels, and trapping
test specify the branches under study, and two independent implementations
of the ingoing/decaying boundary-value problem provide the quantitative checks.

\subsection{{Scope of the charged extension}}

{The calculation in Sec.~\ref{sec:charged} specifies both the
scalar representation and its gauge coupling to the actual Wu--Yang
connection. Coupling to a separate Maxwell field, charged vector or
Dirac perturbations, and additional scalar interactions would require
extensions of this model. The absence of an electric horizon potential and
rotation excludes the usual electrostatic and rotational extraction
channels; the positive-potential argument applies only to the stated
canonical scalar model. Stability of the full coupled EYM system remains a separate problem.}

\subsection{Physical scales and limitations}

If the neutral field represents an ultralight bosonic species of particle
mass \(m_b\), the conversion to physical units is
\begin{align}
 m_bc^2&=1.34\times10^{-10}\,\mathrm{eV}\,
       \muh\frac{M_\odot}{M},\nonumber\\
 t_g&=\frac{GM}{c^3}=4.93\,\mu\mathrm{s}\frac{M}{M_\odot}.
 \label{eq:physicalUnits}
\end{align}
Then \(\tau_{\rm amp}=t_g/\Gam\). A width near \(10^{-10}\)
corresponds to about six days for \(10M_\odot\), or about
\(1.6\times10^3\) years for \(10^6M_\odot\). These examples
keep dimensionless coupling fixed and therefore refer to different
particle masses, illustrating the conversion of spectral widths to physical lifetimes.

Cloud occupation depends on the available field supply and excitation history.
Environmental scalar supply can produce extended configurations around a
black hole \cite{Clough:2019ScalarHair}, but excitation, continued
replenishment, and other depletion channels must be modeled before making
an astrophysical prediction. A macroscopic cloud would eventually require
backreaction. {This means including its stress tensor
(and gauge current for a charged cloud) in the background field equations.
The energy or number of quanta occupying a mode depends on its
excitation and subsequent evolution.} Likewise, a fixed microscopic value of \(\xi\) gives a
very small \(b=\xi/M^2\) for a macroscopic hole; large \(\lambda\)
cases are not automatically observable short-distance corrections.
We make no claim that the exterior test field probes the center or
resolves the nonlinear stability of the inner horizon. Cosmological
horizons, exact extremality, and coupled EYM perturbations are outside
the scope of the calculation.

\section{Conclusions}
\label{sec:conclusion}

While quasinormal spectra have been explored extensively across a wide
range of black-hole configurations
\cite{Bolokhov:2025rng,Konoplya:2006rv,Dubinsky:2024rvf,
Lutfuoglu:2026etg,Lutfuoglu:2026vcr,Konoplya:2023moy,
Konoplya:2005et,Bolokhov:2025egl,Bolokhov:2024otn,
Kodama:2009rq,Skvortsova:2024atk,Skvortsova:2023zmj,
Skvortsova:2026ryl,Lutfuoglu:2026rqe,Lutfuoglu:2026boa,
Lutfuoglu:2026uzy,Konoplya:2017zwo,Lutfuoglu:2026zel,
Lutfuoglu:2025ohb},
the quasi-bound states that support long-lived massive-field clouds
remain comparatively less studied.
{In this work, we have computed quasi-bound states of neutral massive scalars
and canonical charged SU(2) scalar doublets and triplets around the
asymptotically flat nonminimal  EYM black hole. The calculation
separates the geometric charge, nonminimal coupling, and the
probe's quantized gauge sector.}
The principal result is a mass-dependent monopole lifetime reversal.
Nonminimal coupling suppresses absorption in the weak-binding limit,
in agreement with the smaller horizon area, but enhances it beyond
a finite crossover mass relative to the  RN geometry at the
same charge. The dipole and quadrupole reference states instead show
substantial suppression, with changes in their real frequencies far
smaller than the relative changes in their widths.

This conclusion rests on direct complex-frequency calculations, not
on a barrier-height estimate. Two-sided ODE matching and an independent
60-digit continued fraction agree on the reference levels and their
imaginary parts. A complete boundary sweep distinguishes the precision-
qualified modes from ultra-narrow stress tests that are not numerically
resolved to the same relative accuracy. Schwarzschild and  RN
limits, hydrogenic binding, and the leading monopole horizon-area scaling
provide further checks. Several field masses, moderate nonminimal
couplings, and radial excitations establish that the effect is not
confined to one near-extremal example.

The exact horizon-domain analysis also clarifies the zero-charge limit:
Schwarzschild is recovered at finite nonminimal coupling, while holding a
positive nonminimal fraction fixed sends the coupling to infinity.
The potential positivity argument excludes growing neutral test-scalar
modes throughout the nonextremal black-hole domain considered; stability
of the coupled EYM solution remains open. Quasi-bound poles can persist
beyond the disappearance of the potential well.

{The charged calculation replaces the neutral angular eigenvalue
by $j(j+1)-p^2$ through an explicit reduction of the Wu--Yang connection.
Its initial survey contains 304 charged-sector representative poles and 18
neutral-weight controls; all pass the boundary-variation criterion, while
303 additionally satisfy the cloud selection. Twenty charged poles are
independently validated with arbitrary-precision continued fractions.
A static-transfer calculation recovers the analytic  RN limit
and the noninteger-power weak-binding widths.}

{Twelve charged-channel lifetime reversals are located inside
the sampled trapping region. At $e=0.6$, $\lambda=0.9$, the lowest
charged doublet and triplet reverse at $\muh\simeq0.224729$ and
$0.302953$, respectively. A different comparison, at fixed $j=1$
within the triplet, lowers the angular barrier in the charged sector
and enhances its absorption by factors reaching hundreds in the
sampled cases. Gauge charge therefore does not universally protect a
cloud from decay.}

{Continuing the lowest charged branches at $e=0.6$ to
$\muh=0.75$ reveals maxima in their nonminimal/RN width ratios and a
second return to relative suppression. These later crossings lie outside
the three-turning-point cloud region, even though the same ingoing/decaying
poles persist. Their absolute widths continue to grow with mass; the new
ratio reversal therefore does not imply increasingly long-lived clouds.}

{The background remains static and purely magnetic: no electric
Coulomb frequency shift or superradiant growth is introduced. The
positivity argument excludes growing canonical scalar modes but does
not establish nonlinear or coupled EYM stability. Exact extremality,
rotating or dyonic backgrounds, other field spins and interactions, and
cloud population and backreaction remain separate problems.}

\appendix
\section{Boundary series for the ODE method}
\label{app:series}

All quantities are dimensionless. Near \(y=x-x_+\), write
\(f=f_1y+f_2y^2+\cdots\), where \(f_1=f_x(x_+)\) and
\(f_2=f_{xx}(x_+)/2\). With
\(u=y^\rho\sum_{n\geq0}c_ny^n\), \(c_0=1\), the first correction is
\begin{align}
 p_0&=\frac{f_2}{f_1},\qquad q_{-2}=\frac{\om^2}{f_1^2},\\
 q_{-1}&=-2q_{-2}\frac{f_2}{f_1}
       -\frac{\muh^2+\ell(\ell+1)/x_+^2}{f_1}-\frac1{x_+},\\
 c_1&=-\frac{p_0\rho+q_{-1}}{2\rho+1}.
\end{align}
For arbitrary order, multiply Eq.~\eqref{eq:areal} by \(N^2Dx^2\)
and define
\begin{align}
 A_h&=N^2Dx^2,\\
 B_h&=(NN'D-N^2D')x^2,\\
 C_h&=\om^2D^3x^2-\muh^2ND^2x^2
       -\ell(\ell+1)ND^2\nonumber\\
    &\hspace{1em}-(NN'D-N^2D')x.
\end{align}
Primes here act on \(x\). Expand these polynomials in \(y\), with
coefficients \(A_{h,j},B_{h,j},C_{h,j}\). The recurrence is
\begin{align}
 K_{n,m}={}&A_{h,n-m+2}(m+\rho)(m+\rho-1)\nonumber\\
          &+B_{h,n-m+1}(m+\rho)+C_{h,n-m},\\
 c_n={}&-\frac{1}{K_{n,n}}\sum_{m=0}^{n-1}K_{n,m}c_m.
 \label{eq:horizonRecurrence}
\end{align}
Coefficients outside a polynomial's range vanish. The numerical boundary
value is \(u_x/u=\rho/\epsilon+H'(\epsilon)/H(\epsilon)\),
where \(H=\sum_{n=0}^{10}c_ny^n\) {for the baseline
order-ten initialization; the sum extends to the higher order when the
boundary test or high-mass continuation requires it. The outward and
inward series orders are varied together.}

At infinity let
\(u=e^{-kx}x^\nu\sum_{n\geq0}d_nx^{-n}\), \(d_0=1\).
{The radial coefficients have expansions
\(P=2/x^2+O(x^{-3})\) and
\(\mathcal Q=-k^2+2(2\om^2-\muh^2)/x+s_2/x^2+O(x^{-3})\).
Cancelling the \(x^{-1}\) term gives
\(k\nu=2\om^2-\muh^2\); the next order determines \(d_1\).}
The first coefficient is
\begin{align}
 s_2&=\om^2(12-2e^2)-\muh^2(4-e^2)-\ell(\ell+1),\\
 d_1&=-\frac{\nu^2-\nu-2k+s_2}{2k}.
\end{align}
Higher orders follow from a formal large-radius series expansion. In \(v=1/x\), let \(H(v)=\sum d_nv^n\)
and expand \(P\) and \(\mathcal Q\) from Eq.~\eqref{eq:areal}.
Then
\begin{align}
 0={}&v^4H_{vv}+[2kv^2+(2-2\nu)v^3-Pv^2]H_v\nonumber\\
 &+[k^2-2k\nu v+(\nu^2-\nu)v^2
       +P(-k+\nu v)+\mathcal Q]H.
 \label{eq:outerRecurrence}
\end{align}
Equating successive coefficients determines \(d_n\); the coefficient
of the newly introduced \(d_n\) at order \(v^{n+1}\) is \(2kn\).
This construction includes the nonminimal far-field terms at the order
where they first contribute.

\section{Finite-band continued fraction}
\label{app:cf}

Set \(s=1-z\), \(R=x_+-x_-z\), \(\delta=x_+-x_-\), so
\(x=R/s\). The homogeneous metric polynomials are
\begin{align}
 \mathsf N&=R^4-2R^3s+e^2R^2s^2+2be^2s^4,\\
 \mathsf D&=R^4+2be^2s^4.
\end{align}
For \(u=g(z)h(z)\), with the prefactor in Eq.~\eqref{eq:frobenius},
define
\begin{align}
 H&=\frac{g_z}{g}=\frac\rho z+\frac\nu s-\frac{k\delta}{s^2},\\
 \mathsf P&=-\frac2s+\frac{\mathsf N_z}{\mathsf N}
                         -\frac{\mathsf D_z}{\mathsf D},\\
 \mathsf Q&=\left(\frac{dx}{dz}\right)^2\mathcal Q(x(z)),
 \qquad \frac{dx}{dz}=\frac{\delta}{s^2}.
\end{align}
Multiplication by \(A_0=s^4\mathsf N^2\mathsf D R^2\) and cancellation
of the common factor \(zs^2\) yields
\begin{align}
 A&=\frac{A_0}{zs^2},\qquad B=A(2H+\mathsf P),\\
 C&=A(H_z+H^2+\mathsf PH+\mathsf Q),\\
 Ah_{zz}+Bh_z+Ch&=0.
\end{align}
Although written as rational expressions, \(A,B,C\) are polynomials
after the horizon indicial equation and Eq.~\eqref{eq:exponents} are
used. Their generic degrees are 13, 12, and 11. If their coefficients are
\(A_j,B_j,C_j\), the equation for the series coefficients is
\begin{align}
 \sum_{j=0}^{12}c_{n,j}a_{n+1-j}&=0,\\
 c_{n,j}&=A_{j+1}(n+1-j)(n-j)\nonumber\\
        &\quad+B_j(n+1-j)+C_{j-1},
\end{align}
where negative-index \(a_n\) and out-of-range polynomial coefficients
are zero. For a recurrence with last column \(J\), the elimination step is
\begin{equation}
 \widetilde c_{n,j}=c_{n,j}
       -\frac{c_{n,J}}{\widetilde c_{n-1,J-1}}
                         \widetilde c_{n-1,j-1},\quad 1\leq j<J,
\end{equation}
with the leading column unchanged. Repetition gives coefficients
\((\alpha_n,\beta_n,\gamma_n)\) of a three-term recurrence. The
fundamental continued-fraction condition is
\begin{equation}
 \beta_0-\frac{\alpha_0\gamma_1}{\beta_1-
        \displaystyle\frac{\alpha_1\gamma_2}{\beta_2-\cdots}}=0.
\end{equation}
Finite depth sets the tail beyond the last retained coefficient to zero;
its error is tested by increasing that depth. Inverted fractions are used
for initial radial-excitation estimates. The primary ODE calculation and
the precision validation have no shared boundary-series coefficients.
{For completeness, define the finite left denominators by
\begin{equation}
 \mathcal B_0=\beta_0,\qquad
 \mathcal B_n=\beta_n-\frac{\alpha_{n-1}\gamma_n}{\mathcal B_{n-1}},
 \quad 1\leq n\leq n_{\rm inv}.
\end{equation}
The fraction inverted at index \(n_{\rm inv}>0\) equates
\begin{equation}
 \mathcal B_{n_{\rm inv}}=
 \frac{\alpha_{n_{\rm inv}}\gamma_{n_{\rm inv}+1}}{\beta_{n_{\rm inv}+1}-
 \displaystyle\frac{\alpha_{n_{\rm inv}+1}\gamma_{n_{\rm inv}+2}}{\beta_{n_{\rm inv}+2}-\cdots}}.
 \label{eq:inversionReview}
\end{equation}
For \(n_{\rm inv}=1\) the left side is simply
\(\beta_1-\alpha_0\gamma_1/\beta_0\). The inversion
index is set to the desired \(n_r\) for initial estimates, and continuation
then fixes the physical branch. At \(b=0\)
or \(e=0\), further polynomial cancellations can shorten the generic
recurrence. Every appendix formula for the charged scalar uses
\(\ell(\ell+1)\to\Lambda_j\); no other electric term is added.}

\backmatter

\section*{Statements and declarations}
Chat GPT was used for editing and style improvements of this work. 

\subsection*{Data availability}
The data supporting the results are presented explicitly in the manuscript.
Additional data are available from the authors upon reasonable request.

\section*{Acknowledgments}
The author would like to thank Dr. R. A. Konoplya for helpful discussions.

\bibliography{references}
\end{document}